\documentclass[11pt]{article}
\usepackage{amssymb}
\usepackage{amsmath}
\usepackage{fullpage}
\usepackage{color}
\usepackage{accents}
\usepackage{ifpdf}
\ifpdf
   \usepackage[pdftex]{graphicx}
   \usepackage{float}
   \usepackage[caption = false]{subfig}
   \newcommand{\pdf}[1]{#1} \newcommand{\eps}[1]{}
\else
   \usepackage{graphicx}
   \usepackage{float}
   \usepackage[caption = false]{subfig}
   \newcommand{\eps}[1]{#1} \newcommand{\pdf}[1]{}
\fi

\newcommand{\T}{^\mathsf{T}}
\newcommand{\bfa}{\mathbf{a}}
\newcommand{\bfp}{\mathbf{p}}
\newcommand{\bfq}{\mathbf{q}}
\newcommand{\bfQ}{\mathbf{Q}}
\newcommand{\bfu}{\mathbf{u}}
\newcommand{\bfx}{\mathbf{x}}
\newcommand{\bfz}{\mathbf{z}}
\newcommand{\bfZ}{\mathbf{Z}}
\newcommand{\calF}{\mathcal{F}}
\newcommand{\barcalF}{\bar{\mathcal{F}}}
\newcommand{\calN}{\mathcal{N}}
\newcommand{\calO}{\mathcal{O}}
\newcommand{\calR}{\mathcal{R}}
\newcommand{\rmd}{\mathrm{d}}

\newcommand{\rmq}{\mathrm{q}}
\newcommand{\bbE}{\mathbb{E}}
\newcommand{\est}{{\mathrm{est}}}
\newcommand{\Var}{\mathrm{Var}}
\newcommand{\overU}{\overline{U}}

\title{
Quasi-Reliable Estimates of Effective Sample Size
}
\author{
Youhan Fang \\
Purdue University \\
yfang@purdue.edu
\and Yudong Cao \\
Harvard University \\
yudongcao@fas.harvard.edu
\and Robert D. Skeel \\
Purdue University / Arizona State University \\
}

\date{}

\begin{document}

\maketitle

\begin{abstract}
The efficiency of a Markov chain Monte Carlo algorithm
might be measured by the cost of generating one independent sample,
or equivalently, the total cost divided by the effective sample size,
defined in terms of the integrated autocorrelation time.
To ensure the reliability of such an estimate,
it is suggested that there be an adequate sampling of state space---%
to the extent that this can be determined from the available samples.
A possible method for doing this is derived and evaluated.
\end{abstract}

\section{Introduction}

Markov chain Monte Carlo (MCMC) methods are very widely used
to estimate expectations of random variables,
since they are often the only methods
applicable to complicated and high dimensional distributions.
Their accuracy is low relative to computational effort,
so it is particularly important to have reliable error estimates.
Due to the correlated nature of MCMC samples,
a variance estimate for $N$ samples uses the effective sample
size $\mathit{ESS}$ rather than $N$ in the denominator.
The effective sample size for
a random variable
can be computed from its integrated autocorrelation time.
Proposed in this paper
is a more reliable estimate of $\mathit{ESS}$,
particularly for multimodal distributions.

This study
is motivated by an interest in designing more efficient MCMC samplers.
For evaluating performance of MCMC methods, there is wide support
among both statisticians, e.g., Charles Geyer (1992, page 11) \cite{Geye92},
and physicists, e.g., Alan Sokal (1997, page 13) \cite{Soka97}
for using as a metric the cost of an independent sample:
\begin{align*}
\frac{N}{\mathit{ESS}}\times\mbox{cost per step}
 =\tau\times\mbox{cost per step},
\end{align*}
where $\tau$ is the {\em integrated autocorrelation time}.

Similarly, for evaluating accuracy of MCMC estimates,
it is deemed essential, e.g., Alan Sokal (page 11) \cite{Soka97} again,
to know the autocorrelation time.
Nonetheless, MCMC is often used for problems for which $\mathit{ESS}$ is
very small, perhaps even less than 1.
How should such problems be handled?
One option is to avoid claims of sampling and instead designate
the simulation as ``exploration''.
Another option is to
change the problem:
For example, use a simpler model.
A coarsened model with error bars is better
than  a refined model without error bars
(because then the limitations are more explicit).
Alternatively, artificially limit the extent of state space,
and adjust conclusions accordingly.

A computed estimate of the integrated autocorrelation time
might be marred by incomplete sampling---at
least in the case of multimodal distributions.
Without additional information,
there is no foolproof way of detecting convergence,
so reliability is impractical in general.
So the goal is instead quasi-reliability,
which is defined to mean
apparent good coverage of state space---%
ensuring thorough sampling of
that part of state space that has already been explored,
to minimize the risk of missing an opening
to an unexplored part of state space.
More concretely, it might mean
thorough sampling of 
those modes that have already been visited,
to minimize the risk of missing an opening to yet another mode.

This article makes a couple of contributions towards more
reliable estimates of effective sample size:
\begin{enumerate}
\item A quantity is derived for measuring thoroughness of sampling
that is simple to describe and relatively easy to estimate, namely,
the longest autocorrelation time over all possible 
functions defined on state space,
denoted by $\tau_{\max}$.
This quantity has two advantages over estimates of $\tau$
only for functions of interest.
(i) It is more likely to indicate the existence of a transition path
to a mode that has not been visited.
(ii) It is more reliably estimated with finite sampling
than is $\tau$ for just any function:
even though a function of interest might have a small value of $\tau$,
a reliable estimate of
its autocorrelation function and integrated autocorrelation time
$\tau$ may well require a number of samples
dictated by $\tau_{\max}$.
\item An algorithm is constructed for estimating $\tau_{\max}$,
 assuming the availability of a method for estimating
 the integrated autocorrelation time $\tau$ for an arbitrary function,
 and evidence is presented for the utility of this algorithm.
\end{enumerate}

Section~2 of this article establishes notation by reviewing
needed concepts and methods.
It presents the definition and role of the integrated autocorrelation $\tau$.
It defines a {\em forward transfer operator}
that comprises transition probabilites of the MCMC process.
And it presents the notion of a lag window for estimating $\tau$
for a finite number of samples $N$.

Section~3 is the heart of this article.
It begins with an example illustrating how it is possible to have an
apparently reliable esimate of $\tau$ that is far too small.
It then defines thorough sampling to mean that (with, say, 95\% confidence)
the fraction of samples from any given subset of state space
should differ from the actual probability
by no more than some specified tolerance $\mathit{tol}$.
It is shown that this is achieved if $N\ge\tau_{\max}/\mathit{tol}^2$.
An approximate maximum is obtained by considering arbitrary
linear combinations of a finite set of basis functions.
This results in a generalized eigenvalue problem.
The matrices are shown to be symmetric not only for reversible MCMC methods
but also for a class of ``modified''  reversible MCMC methods
that includes generalized hybrid Monte Carlo samplers
and reversible (inertial) Langevin integrators.

Section~4 tests the methodology on 4 examples.
A 1-dimensional Gaussian distribution is used to confirm the correctness of the theory.
A sum of 2 Gaussians is used as a simple example of a multimodal problem,
which demonstrates the advantage of seeking the longest autocorrelation time.
A 2-layer single-node neural network is used to capture the flavor
of machine learning applications.
It yields a large enough estimate of $\tau_{\max}$ to serve as a warning
that an apparently large sample size is inadequate,
unlike the $\tau$ estimate for the quantity of interest.
Finally a logistic regression example confirms again the wisdom
of seeking the longest autocorrelation time $\tau_{\max}$.

Section~5 is a brief discussion and conclusion.

Initial attempts to estimate the integrated autocorrelation time employed
a small computer program called {\tt acor}
\cite{Good09}.
Due to concerns about its reliability, Appendix~\ref{app:tau} derives
an alternative method for defining the lag window under
a couple of simplyfying assumptions, one of which becomes more credible
for observables having a large value of $\tau$.
Experiments indicate slightly more reliable estimates compared to {\tt acor}.

\section{Preliminaries and Background}

Given a probability density $\rho_\rmq(\bfq)$,
$\bfq\in\mathbb{R}^d$,
known only up to a multiplicative factor,
the aim is to compute observables $\bbE[u(\bfQ)]$,
which are expectations for specified functions $u(\bfq)$.
{\it Note} here the use of upper case for random variables.

Consider a Markov chain
$\bfQ_0\rightarrow \bfQ_1\rightarrow\cdots\rightarrow \bfQ_{N-1}$
that is ergodic with stationary density $\rho_\rmq(\bfq)$.
Also, assume $\bfQ_0\sim\rho_\rmq(\bfq)$.
To estimate an observable, use
\begin{align*}
\bbE[u(\bfQ)]\approx\overU =\frac1{N}\sum_{n=0}^{N-1} u(\bfQ_n),
\end{align*}
but use just one realization
$\bfq_0\rightarrow \bfq_1\rightarrow\cdots\rightarrow \bfq_{N-1}$,
assuming samples are renumbered to omit those prior to equilibration/burnin.

\subsection{Variance of the estimated mean}

There are
a variety of methods for estimating the variance of the estimated mean.
Most appealing~\cite{Geye92} are those that expoit the fact that
the sample is generated by a Markov chain.
The variance of the estimated mean is given by
\begin{align*}
\Var[\overU] =\frac1{N}\Var[u(\bfQ)]
\left(1 + 2\sum_{k=1}^{N-1}\left(1 -\frac{k}{N}\right)\frac{C(k)}{C(0)}\right)
\end{align*}
where the autocovariances
\begin{align*}
C(k) =\bbE[(u(\bfQ_0) -\mu)(u(\bfQ_k) -\mu)],
\end{align*}
with $\mu =\bbE[u(\bfQ)]$. In the limit $N\rightarrow\infty$,
\[\Var[\overU] =\frac1{N}\Var[u(\bfQ)]\tau + o(\frac1{N})\]
where $\tau$ is the {\it integrated autocorrelation time}.
\begin{equation}  \label{eq:tau}
\tau = 1 + 2\sum_{k=1}^{+\infty}\frac{C(k)}{C(0)}.
\end{equation}

An estimate of  covariances is provided by
Ref.~\cite[pp.~323--324]{Prie81}:
\begin{align*}
C_N(k) =\frac1{N}
 \sum_{n=0}^{N-k-1}(u(\bfq_n) -\overline{u})(u(\bfq_{n+k}) -\overline{u}).
 \end{align*}

\subsection{Forward transfer operator}

Some MCMC samplers
augment state variables $\bfq$ with auxiliary variables $\bfp$, e.g., momenta,
extend the p.d.f. to $\rho(\bfq,\bfp)$ so that
\begin{align*}
 \int\rho(\bfq,\bfp)\rmd\bfp =\rho_\rmq(\bfq),
 \end{align*}
and make moves in extended state space
$\bfz_0\rightarrow\bfz_1\rightarrow\cdots\rightarrow\bfz_{N-1}$
where $\bfz = (\bfq,\bfp)$.

Associated with the MCMC propagator is an operator $\calF$,
which maps a relative density $u_{n-1} =\rho_{n-1}/\rho$ for $\bfZ_{n-1}$
to a relative density $u_n =\rho_n/\rho$ for $\bfZ_n$:
\begin{align*}
u_n =\calF u_{n-1}
\end{align*}
where
\begin{align*}
\calF u_{n-1}(\bfz)
 =\frac1{\rho(\bfz)}\int\rho(\bfz|\bfz')u_{n-1}(\bfz')\rho(\bfz')\rmd\bfz'
\end{align*}
and $\rho(\bfz|\bfz')$ is the transition probability density for the chain~\cite{sch00}.
Note
that $u\equiv 1$ is an eigenfunction of $\calF$ for eigenvalue $\lambda_1 =1$.

The error in the probability density,
starting from $\rho_0(\bfq) =\delta(\bfq -\bfq_0)$, is
\begin{align*}
\frac1{N}\sum_{n=0}^{N-1}\rho_n -\rho
 =\rho\frac1{N}(1 -\calF_0)^{-1}(1 -\calF_0^N)\frac{\rho_0 -\rho}{\rho}
\end{align*}
where $\calF_0$ is $\calF$ with the eigenvalue 1 removed,
i.e., $\calF_0 u =\calF u -\bbE[u(\bfQ)]$.
From this, one sees that the error is proportional to the reciprocal
of the spectral gap $|1 -\lambda_2|$, where $\lambda_2$ is the
nonunit eigenvalue of $\calF$ nearest 1.
This, however, is not the relevant quantity---in general (see \ref{subsec:thor}).

\subsection{Error in estimates of integrated autocorrelation time}

The standard deviation of the statistical error in the estimate
of $\tau$ from Eq.~(\ref{eq:tau}) grows with the number of terms taken;
more specifically, it is approximately
$\sqrt{M/N}\tau$
where $M$ is the number of terms~\cite[Eqs.~(3.19)]{Soka97}.
Therefore, one uses instead
\begin{align*}
\tau\approx 1 + 2\sum_{k=1}^{N-1}w(k)
\frac{C_N(k)}{C_N(0)}
\end{align*}
where $w(k)$ is a decreasing weight function  known as a {\em lag window}.

The tiny program {\tt acor} uses a lag window
that is $1$ for $k$ from $0$ to $M-1$ and $0$ for the rest
where $M$ is the smallest number that exceeds the estimated
$\tau$ by a factor  of 10.
It requires the number of samples $N$ to exceed $100\tau$.

\section{Thorough Sampling}
\label{sec:thor}

Depending on estimates of the ESS for just the observables of interest can be deceptive.

\begin{figure*}
\label{fig:2dgauss_samples}
  \centering 
  \subfloat[5000 samples]{\includegraphics[width=0.375\textwidth]{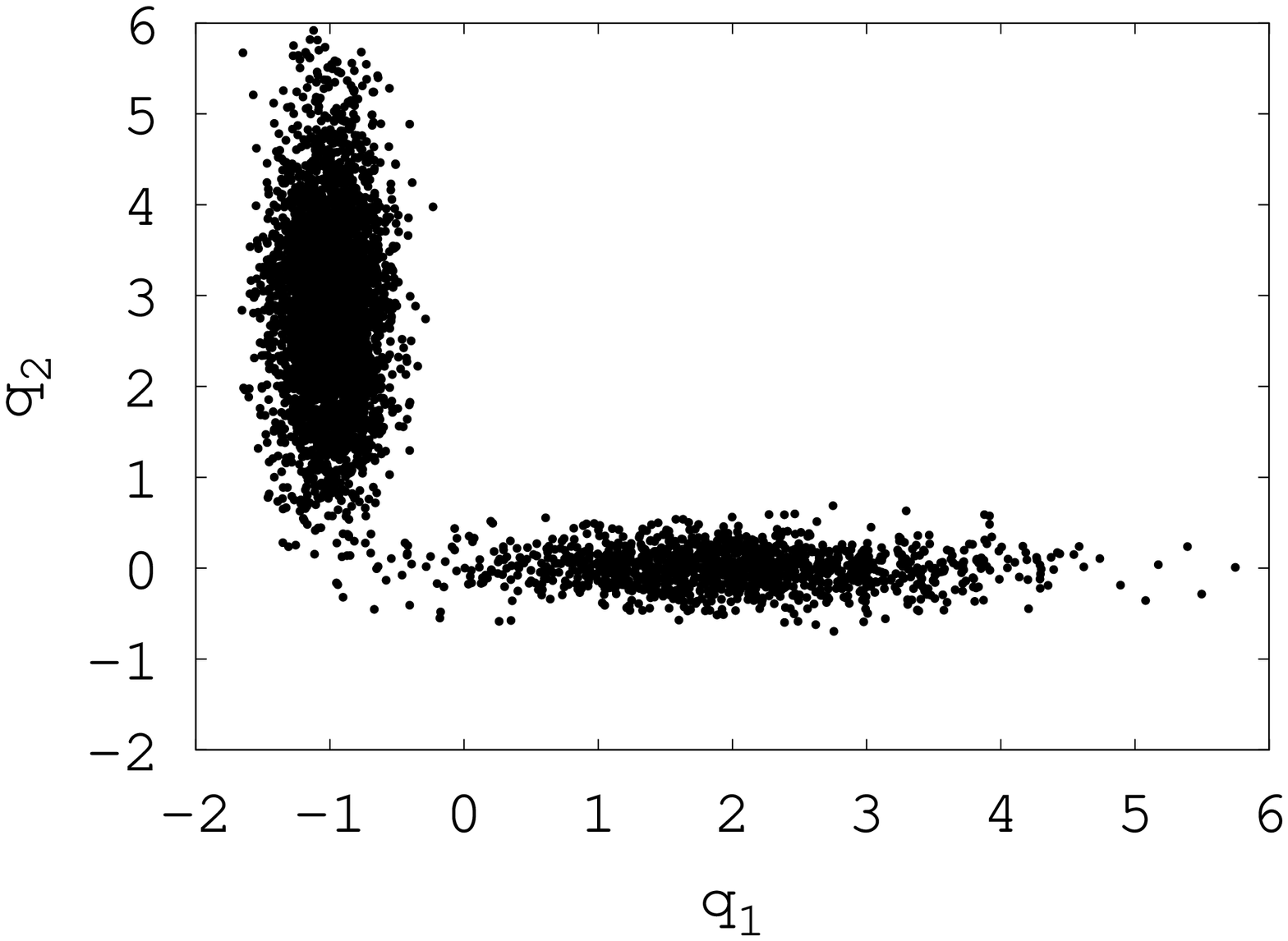}}
  \subfloat[Trajectories of $q_1$ and $q_2$
  ]{\includegraphics[width=0.375\textwidth]{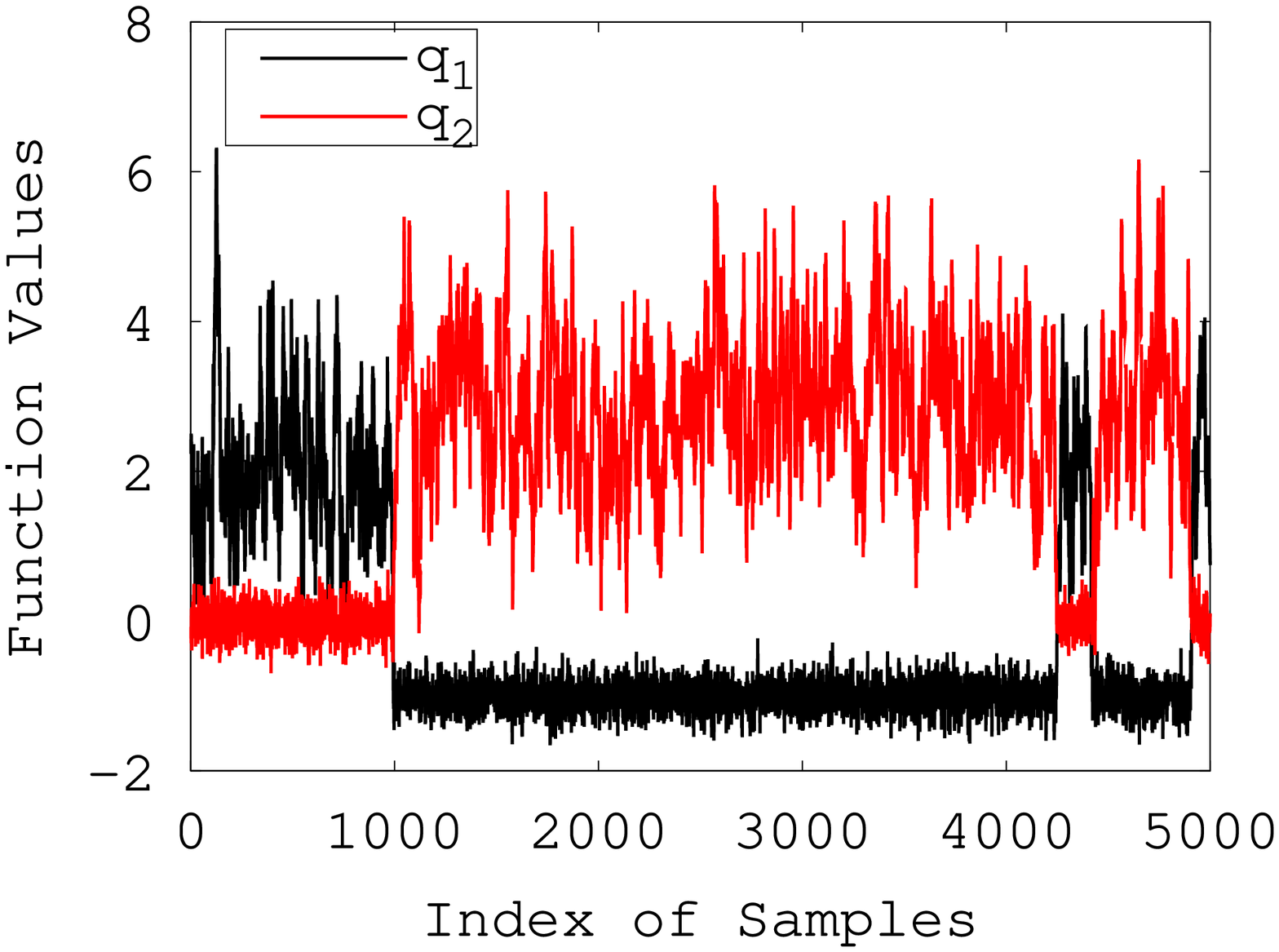}}   
  \caption{
  Example trajectory for the L-shaped mixture of Gaussians distribution.
   }
  
\end{figure*}

Consider a mixture of 2 Gaussians,
\begin{align}\label{eq:2dgauss}
-\log\rho(q_1, q_2) =&\frac12(36(q_1+1)^2 + (q_2-3)^2)\\\nonumber
                      &+\frac12((q_1-2)^2 + 36q_2^2) +\mbox{const},
\end{align}
whose combined basin has an L shape
with a barrier at the corner of the L.
For a sampler, consider the use of Euler-Maruyama Brownian dynamics (w/o rejections),
defined in Sec.~\ref{ss:symmetry}.
A realization of  a sampling trajectory is  given in
Figure \ref{fig:2dgauss_samples}.
For the realization shown there, 
the trajectory makes the first transition around $N = 1000$.

\begin{figure*}
  \centering 
  \subfloat[Autocorrelation time]{\includegraphics[width=0.375\textwidth]{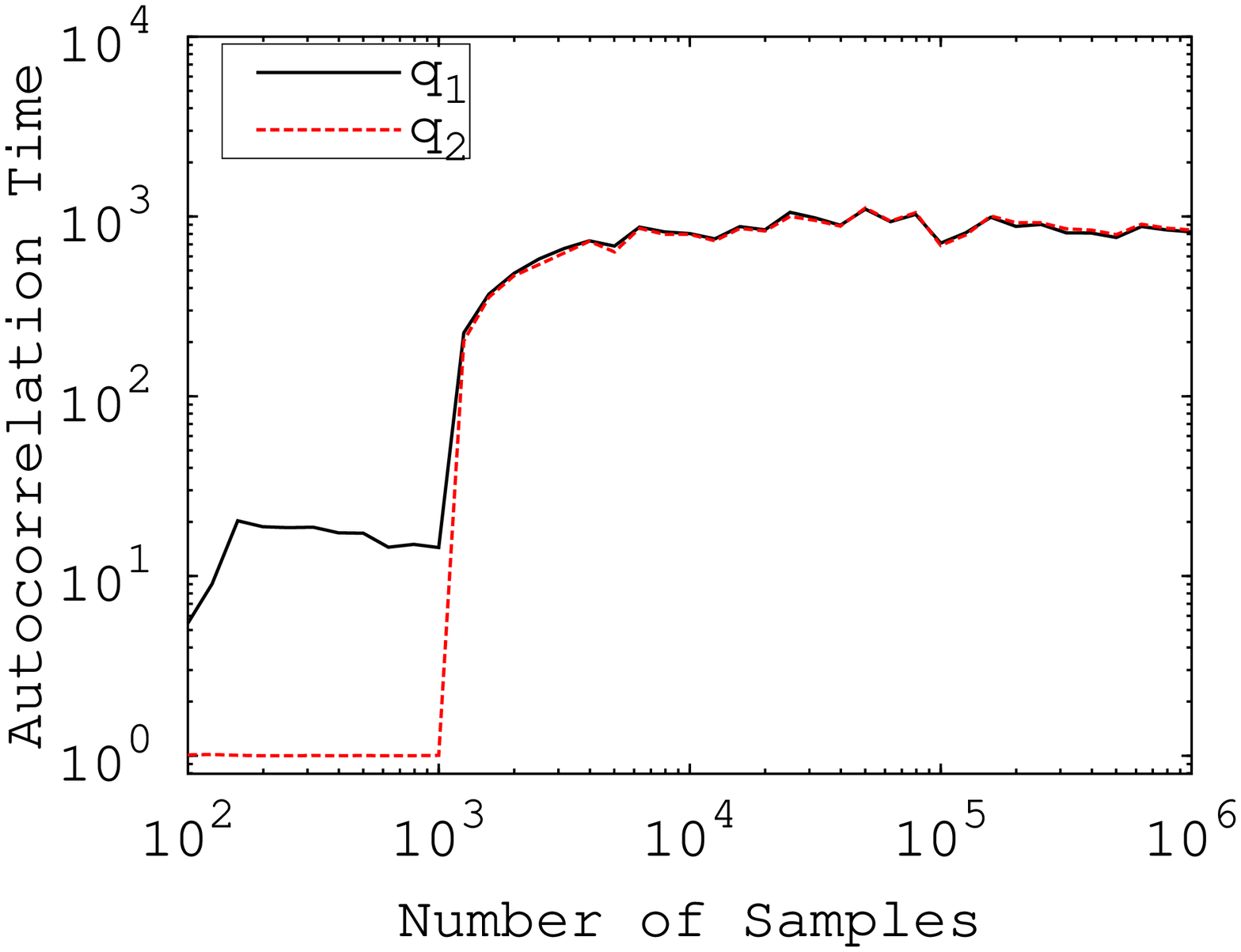}}
  \subfloat[Effective sample size]
  {\includegraphics[width=0.375\textwidth]{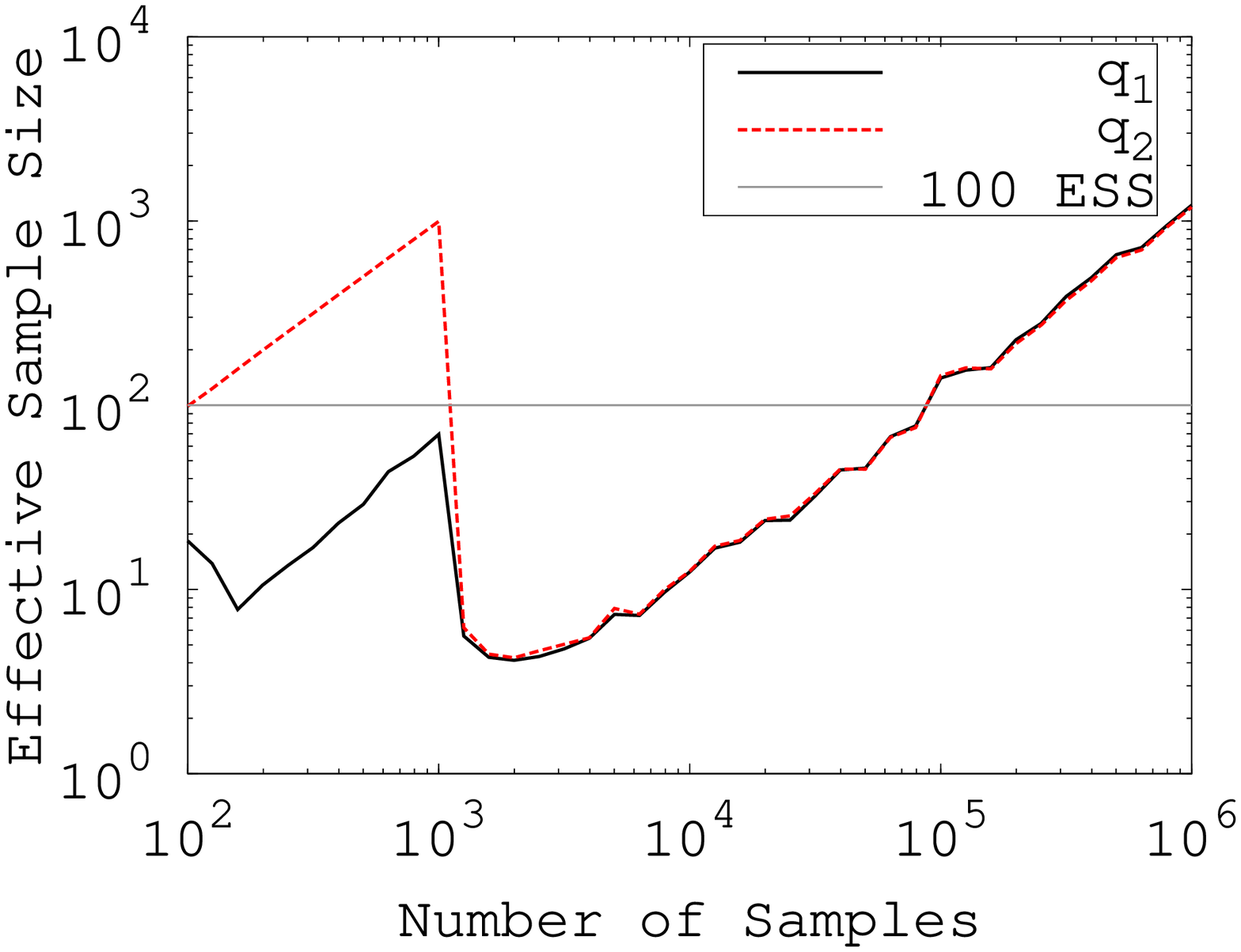}}   
  \caption{
  The autocorrelation time and the effective sample size vs. the number of samples in
  the L-shaped mixture of Gaussians problem.
   }
  \label{fig:2dgauss_tau12}
\end{figure*}

Figure \ref{fig:2dgauss_tau12} shows the estimated integrated autocorrelation time and the effective sample size
as a function of the sample size.
To consider the question of thorough sampling,
suppose that $\mathit{ESS} = 100$ is regarded adequate
for estimating observables.
Just before the $q_1$ curve jumps,
the ESS is much greater than 100 indicating thorough sampling.
This is misleading since after the jump the ESS drops dramatically.
On the other hand,
the $q_2$ curve more reliably detects thorough sampling. Note that if the trajectory
starts from the vertical leg of the "L", the behavior of $q_1$ and $q_2$ are swapped.
This implies that a general observable that is independent of the starting point and can
capture the ``hardest'' move is needed for better detecting convergence. 

\subsection{Definition of thorough sampling}
\label{subsec:thor}

Good coverage of state space might be defined as follows:
for any subset $A$ of (unextended) state space,
an estimate $\overline{1_A}$ of $\bbE[1_A(\bfQ)] =\Pr(\bfQ\in A)$
satisfies
\[\Var[\overline{1_A}]\le\frac14\mathit{tol}^2.\]
This would ensure 95\% confidence
that the proportion of samples for any subset $A$
is not off by more than $\mathit{tol}$.
Since
\[\Var[\overline{1_A}]
\approx\tau_A\frac1{N}\Var[1_A(\bfQ)]\le\frac1{4N}\tau_A,\]
it is enough to have
\begin{equation}  \label{eq:thor}
\frac1{4N}\tau_A\le\frac14\mathit{tol}^2\quad\mbox{for all }A.
\end{equation}

There may be permutations $P$ such that
$\rho_\mathrm{q}(P\bfq) =\rho_\mathrm{q}(\bfq)$ and
$u(P\bfq) = u(\bfq)$ for all interesting $u$.
Then, it is appropriate to weaken the definition of good coverage
by including only those $A$ for which $1_A(P\bfq) = 1_A(\bfq)$.

To facilitate analysis,
introduce the inner product
\[\langle v, u\rangle = \int\overline{v(\bfz)}u(\bfz)\rho(\bfz)\,\rmd\bfz.\]
Note
that $\langle 1, u\rangle$ is the expectation $\bbE[u(\bfZ)]$ of $u(\bfZ)$.
One can show by induction that the cross covariance
\[
\bbE[v(\bfZ_0)u(\bfZ_k)] =\langle\calF^k v, u\rangle.
\]
and, in particular, $C(k) =\langle\calF^k u, u\rangle$.

For simplicity, instead of just indicator functions $1_A(\bfq)$,
consider arbitrary functions in
\begin{align*}
W =&\{u = u(\bfq)~|~
 \bbE[u(\bfQ)] = 0, \\\nonumber
 &u(P\bfq) = u(\bfq)\mbox{ for symmetries }P\},
\end{align*}
and define thorough sampling using
\[\tau_{\max} =\max_{u\in W}\left(1
 + 2\sum_{k=1}^{+\infty}\frac{C(k)}{C(0)}
\right)\]
in Eq.~(\ref{eq:thor}).
The maximum autocorrelation time can be rewritten as
\begin{align*}
\tau_{\max}
 =&\max_{u\in W}\left(1
 + 2\sum_{k=1}^{+\infty}\frac{\langle\calF^k u, u\rangle}{\langle u, u\rangle}
\right)\\
 =&\max_{u\in W}
 \frac{\langle g(\calF) u, u\rangle}{\langle u, u\rangle}
\end{align*}
where
\[g(\lambda) = 1 + 2\lambda/(1 -\lambda).\]

For reversible samplers, the spectral gap is intimately related to $\tau_{\max}$.
If $\tau_{\max}$ were the maximum over {\em all} $u(\bfz)$, then
\[\tau_{\max} =\frac{1 +\lambda_2}{1 -\lambda_2}.\]
where $\lambda_2$ is the second largest eigenvalue of the
transfer operator.

\subsection{Spatial discretization}

For computational purposes,
the idea is to seek a function that comes close to maximizing
the autocorrelation time by considering a linear combination
$u(\bfq) =\bfa\T\bfu(\bfq)$ of given basis functions $u_i\in W$
where the $a_i$ are to be determined.
(In practice, only the \emph{estimated} mean of the function $u(\bfq)$ vanishes.)
Then its autocovariance
\[C(k) =\langle\calF^k u, u\rangle =\bfa\T C_k\bfa\]
where
\[ C_k = \langle\calF^k\bfu, \bfu\T\rangle =\bbE[\bfu(\bfQ_0)\bfu(\bfQ_k)\T]\]
and
\[\tau_{\max}\approx\max_\bfa\frac{\bfa\T K\bfa}{\bfa\T C_0\bfa}\quad
\mbox{where }K = C_0 + 2\sum_{k=1}^{+\infty}C_k\]
---a generalized eigenvalue problem
(arising from maximizing a Rayleigh quotient).

Linear functions $u_i(\bfq) = q_i$ are suggested as a general choice.

\subsection{Symmetry of cross covariance matrices}  \label{ss:symmetry}

The matrix $C_0$ is symmetric positive definite,
if the components of $\bfu$ are linearly independent.
The eigenvalue problem is well conditioned if $K$ is a symmetric matrix. 
Symmetry of $K$ does not hold for every sampler.
Even when $K$ is symmetric, its  finitely sampled estimate is not,
so using the symmetric part can reduce sampling error.

A reversible MCMC samplers is defined to be one that satisfies detailed balance:
\[ \rho(\bfz'|\bfz)\,\rho(\bfz) =\rho(\bfz|\bfz')\,\rho(\bfz')\]
Examples of reversible samplers include
\begin{enumerate}
\item  a Brownian sampler, the Euler-Maruyama discretization of Brownian dynamics
 coupled with a
 Metropolis-Rosenbluth-Rosenbluth-Teller-Teller ~\cite{MRRT53}
 acceptance test
 (known as MALA in the statistics literature~\cite{Rob96}),
\item hybrid Monte Carlo ~\cite{DKPR87},
\item generalized hybrid Monte Carlo~\cite{Horo91},
 with each step followed by a momentum flip,
\item a reversible Langevin integrator,
 with every fixed number of steps followed by a momentum flip (see below).
 Note that reversible has a different meaning for integrators
 than it does for samplers.
\end{enumerate}
These last two samplers augment state space with momentum variables.

The momentum flip of the last two enumerated samplers defeats the purpose
of augmenting state space with momenta.
Indeed, reversing the momentum flip leads to more rapid mixing.
More generally, a {\em modified} reversible propagator effectively
couples the reversible substep $\bfz_{n-1}\rightarrow\bar{\bfz}_n$
with a substep,
\[ \bfz_n = R(\bar{\bfz}_n),\]
where $R$ satisfies $R\circ R =\mathrm{id}$ and $\rho\circ R = \rho$.
For a momentum flip, $R(\bfq,\bfp) = (\bfq, -\bfp)$.

Following is an example of a reversible  Langevin integrator.
Let $F(\bfq) = -\nabla_\bfq\log\rho(\bfq,\bfp)$.
Then each step of the integrator consists of the following
five stages:
\begin{itemize}
\item[B:] $\bfp :=\bfp +\frac12\Delta t F(\bfq)$
\item[A:] $\bfq :=\bfq +\frac12\Delta t\bfp$
\item[O:] $\bfp
 :=\sqrt{1 - 2\gamma\Delta t}\bfp +\sqrt{2\gamma\Delta t}\calN(0, I)$
\item[A:] $\bfq :=\bfq +\frac12\Delta t\bfp$
\item[B:] $\bfp :=\bfp +\frac12\Delta t F(\bfq)$
\end{itemize}
The dynamics of this integrator has a {\em precise} stationary density.
It differs from the desired density~\cite{LeMS16} by $\calO(\Delta t^2)$.
Such error is much smaller than statistical error.
The Euler-Leimkuhler-Matthews integrator~\cite{Leim13}
is the special case $\gamma = 1/(2\Delta t)$ of this,
and it has the remarkable property of retaining second order accuracy.
Since applying a momentum flip has no effect in this case,
it is a reversible MCMC propagator.
(The Euler-Leimkuhler-Matthews integrator can be expressed as
a Markov chain in $\bfq$-space if
post-processing is employed~\cite{Vilm15}.)

For a modified reversible propagator,
the forward transfer operator is a product
\[\calF =\calR\barcalF,\]
where $\barcalF$ is the forward transfer operator of the reversible substep
and $\calR$ is the operator for the added substep $\bfz_n = R(\bar{\bfz}_n)$:
\[u_n =\calR\bar{u}_{n-1}\]
where
\[\calR\bar{u}_{n-1}(\bfz) =\frac1{\rho(\bfz)}
 \int\delta(\bfz - R(\bfz'))\bar{u}_{n-1}(\bfz')\rho(\bfz')\rmd\bfz'.\]
It can be shown that the operators $\barcalF$ and $\calR$ are self adjoint:
\[\langle\barcalF g, f\rangle =\langle g,\barcalF f\rangle\quad
\mbox{and}\quad\langle\calR g, f\rangle =\langle g,\calR f\rangle.\]
{\em Additionally}, assume $\calR\bfu(\bfq) =\bfu(\bfq)$.
Then it is straightforward to show that matrices $C_k$ are symmetric.

\subsection{A method for estimating maximum autocorrelation time}

Begin with a guess for $\bfa$, e.g.,
choose $\bfa$ to be a vector of zeros except for a one corresponding to the observable
with the the longest autocorrelation time.
The following algorithm might converge:
\begin{enumerate}
\item
 With $c(k) = \bfa\T C(k)\bfa$,
 use the method of Sec.~\ref{ss:lag} to select $w(k)$, $k = 1, 2,\ldots, M-1$,
where $M$ is heuristically chosen to be large
but not so large to impact computational efficiency,
\item Set
 \[K = C_0 + 2\sum_{k=1}^{M-1}w(k)C_k.\]
\item Choose $\bfa$ to maximize $\bfa\T K\bfa/(\bfa\T C_0\bfa)$.
\item Repeat.
\end{enumerate}

The number of samples $N$ needed for an estimate of $\tau_{\max}$ depends
on $\tau_{\max}$ itself; 
for this one can use a formula suggested by others,
e.g., {\tt acor} requires $N\ge 100\tau$, roughly.

To improve the convergence of the algorithm,
one might prohibit $\tau_{\max}$ from decreasing from one iteration
to the next.

It is important to keep the method simple, i.e., the time complexity must not exceed the complexity for sampling.  And keeping this modest is aided by
use of the FFT to estimate autocovariances.

\section{Numerical Experiments}

\subsection{A Gaussian distribution}

To confirm the correctness of the
theory, consider a simple test problem for sampling where the target distribution 
is the standard Gaussian. The sampler is Brownian dynamics with a discrete time step of length
$\Delta t = 0.02$.
This step size is much smaller 
than needed in practice. It is chosen in this way to make the discretization error negligible,
thereby permitting the use of analytical results
derived for exact Brownian dynamics.
The Markov chain is obtained by
subsampling the original trajectory at intervals of $0.1$ (every 5th point).
This gives
the true eigenvalues of the transfer operator~\cite{Risk96} as $1, \exp(-0.1),\exp(-0.2),\exp(-0.3),\ldots$.
The corresponding eigenfunctions are the Hermite polynomials
$H_i(q) = 2qH_{i-1}(q)-H'_{i-1}(q)$ with $H_0(q) = 1$.

Consider three observables
\[H_3+H_2+H_1,\quad H_3-H_2+H_1,\quad -H_3+H_2+H_1.\]
The goal of this test is to recover the theoretical value
\[\tau_{\max}=(1+\exp(-0.1))/(1-\exp(-0.1))\approx 20.0,\]
and its corresponding observable, which is $H_1(q)$, by using the linear 
combination of given observables.
Twelve independent runs are performed to evaluate
the reliability of the method.

\begin{figure*}
  \centering
  \includegraphics[width=0.75\textwidth]{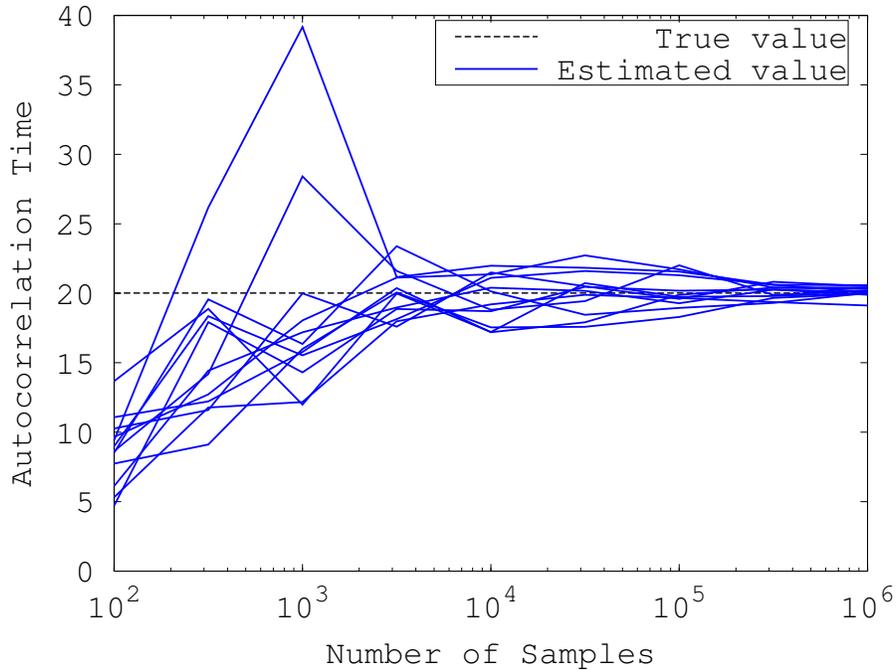}
  \caption{The estimated maximum autocorrelation time in the 1D Gaussian problem.}
  \label{fig:1dgauss_tauall}
\end{figure*}

Figure \ref{fig:1dgauss_tauall}
shows the estimated $\tau_{\max}$ for all 12 runs for increasing sample size. 
It can be seen that the estimated value converges to the true value.
And the variance is comparatively small when $N>3\times 10^3.$
Table \ref{table:1dgauss_x} shows the coefficients
of the linear combinations of given observables.
It can be seen that the theoretical maximizing observable $H_1(q)$ is successfully recovered.

\begin{table}
  \centering
  \caption{The weights of the linear combination of the three observables
   with $a_1$ normalized to 1.}
  \begin{tabular}{lllll}
    \hline\noalign{\smallskip}
       & $N= 10^3$& $N= 10^4$ & $N= 10^5$ & $N= 10^6$ \\ 
       \hline\noalign{\smallskip}
    $a_1$ & -0.036 & -0.008 & 0.003 & 0.010 \\ 
    $a_2$ & 0.972 & 1.024 & 1.008 & 0.994 \\ 
    $a_3$ & 1.000 & 1.000 & 1.000 & 1.000 \\ 
    \hline\noalign{\smallskip}
  \end{tabular}
   
   \label{table:1dgauss_x}
\end{table}


\subsection{An L-shaped mixture of two Gaussian distributions}

Consider again the mixture of two Gaussians with target density given
by eqn.~(\ref{eq:2dgauss}) sampled with discrete Brownian dynamics with time 
step size $\Delta t = 0.02$ and subsampled 
at time interval $0.1$ (every 5 points). The given observables are $q_1$ and $q_2$.
Theoretically, $q_1$ should have larger autocorrelation time than 
$q_2$ for small $N$ while the chain remains in the horizontal leg of the "L", 
since the frequency is smaller in the direction of $q_1$ than it is in the direction
of $q_2$. By the same reasoning, the weight for $q_1$ in the linear combination 
forming the maximizing observable should be greater than the weight for
$q_2$, when $N$ is small. Due to symmetry, on the other hand, both
auto-correlation times and the magnitudes of the weight for 
$q_1$ and $q_2$ should be equal when $N$ large.

\begin{figure*}
  \centering 
  \subfloat[Auto-correlation time]{\includegraphics[width=0.375\textwidth]{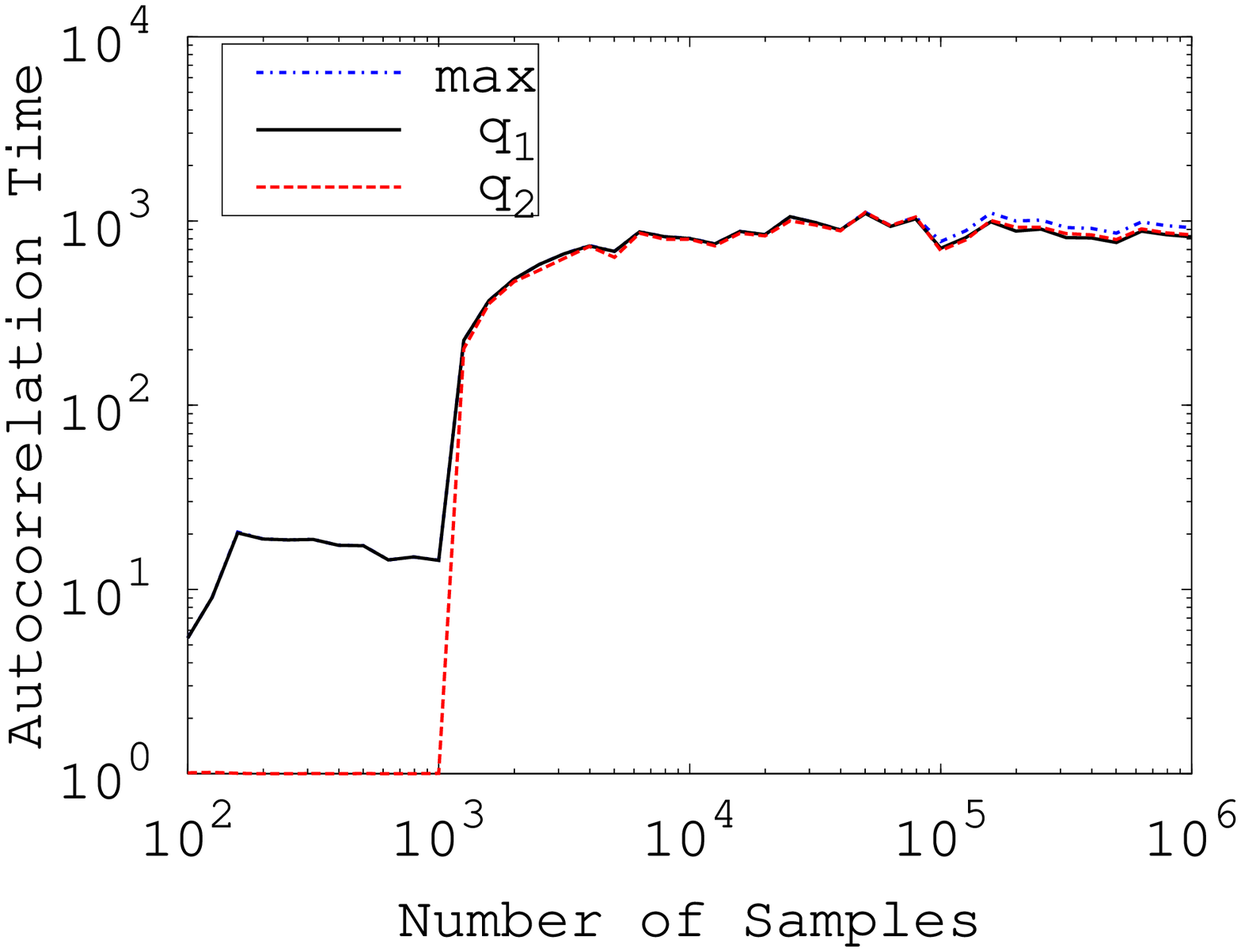}}
  \subfloat[Effective sample size]{\includegraphics[width=0.375\textwidth]{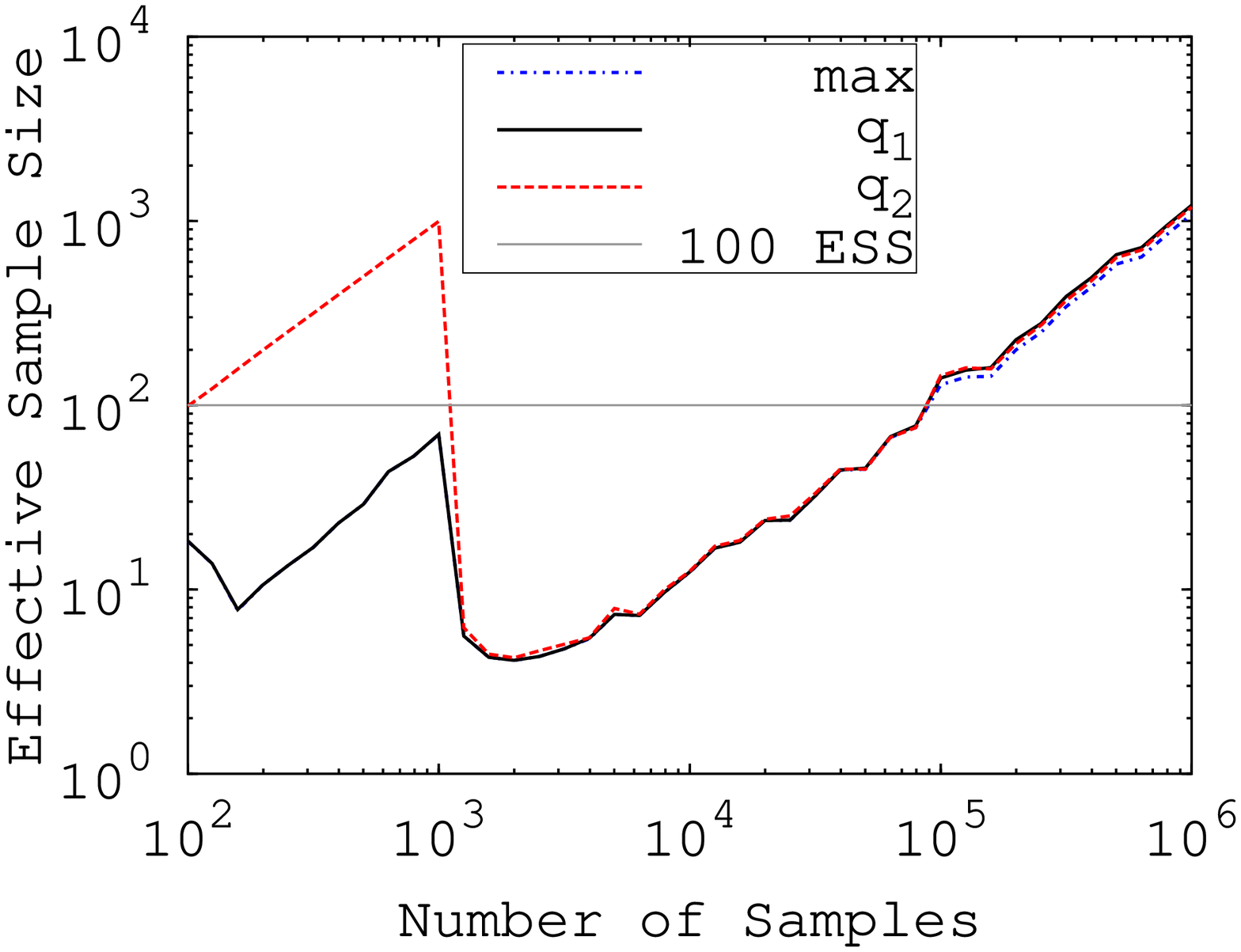}}   
  \caption{
  Effective sample size and autocorrelation time vs.\ number of samples
  in the L-shape mixture of Gaussians.
 }
  \label{fig:2dgauss_ESS_tau}
\end{figure*}

Figure \ref{fig:2dgauss_ESS_tau}
shows ESS and $\tau$ of single observables and the maximizing
observable. It can be seen that $\tau_{\max}$ is always greater than the larger
$\tau$ of individual observables. As is discussed in previous section, 
this tells us that obtaining $\tau_{\max}$
is helpful for better convergence detection. Table \ref{table:2dgauss_x}
shows the weights of linear combinations
for different sample sizes. The weighting obtained meets expectations,
ultimately giving equal weight to $q_1$ and $q_2$.


\begin{table}
  \centering
  \caption{The  weights
    of the linear combination of the $q_1$ and $q_2$ with $a_1$ normalized to $1$.}
  \begin{tabular}{lllll}
    \hline\noalign{\smallskip}
       & $N= 10^3$& $N= 10^4$ & $N= 10^5$ & $N= 10^6$ \\ 
       \hline\noalign{\smallskip}
    $a_1$ & 1.000 & 1.000 & 1.000 & 1.000 \\ 
    $a_2$ & 0.089 & 0.085 & -1.207 & -1.051 \\ 
    \hline\noalign{\smallskip}
  \end{tabular}
   
   \label{table:2dgauss_x}
\end{table}

\subsection{A one-node neural network}

\begin{figure*}
  \centering
  \subfloat[The prediction]{\includegraphics[width=0.3\textwidth]{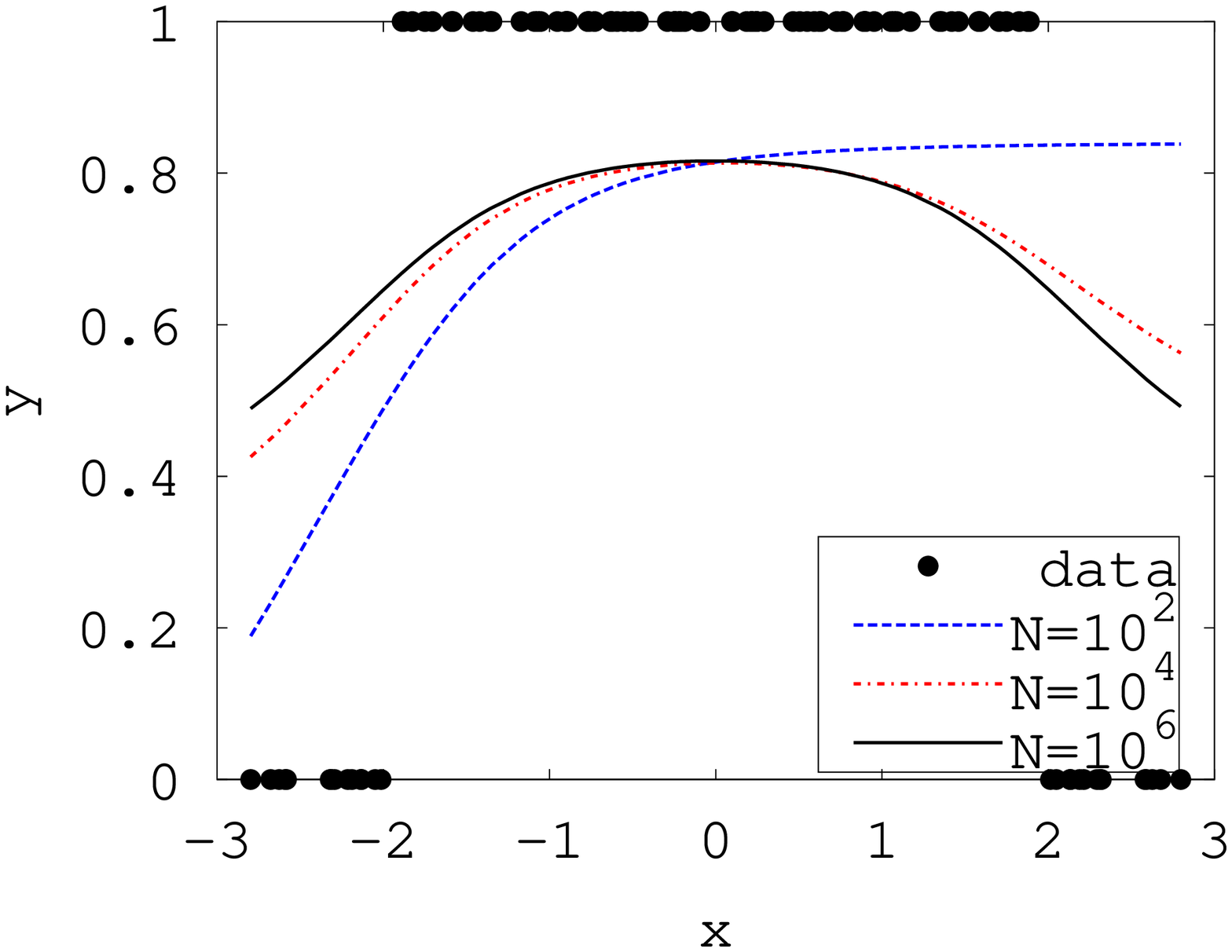}}
  \subfloat[Trajectory of $q_1$]{\includegraphics[width=0.3\textwidth]{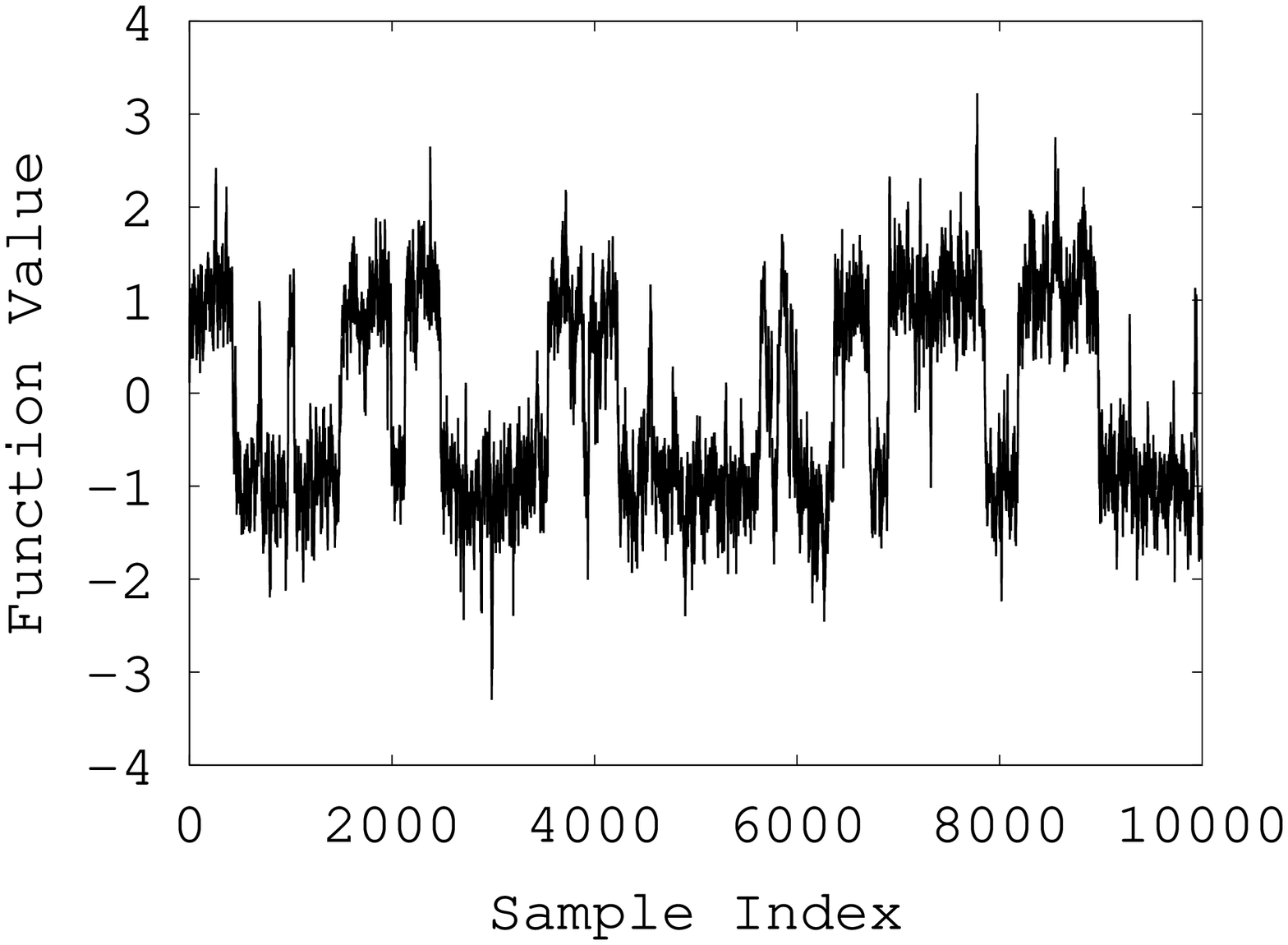}}
  \subfloat[Trajectory of $u_{\max}$]{\includegraphics[width=0.3\textwidth]{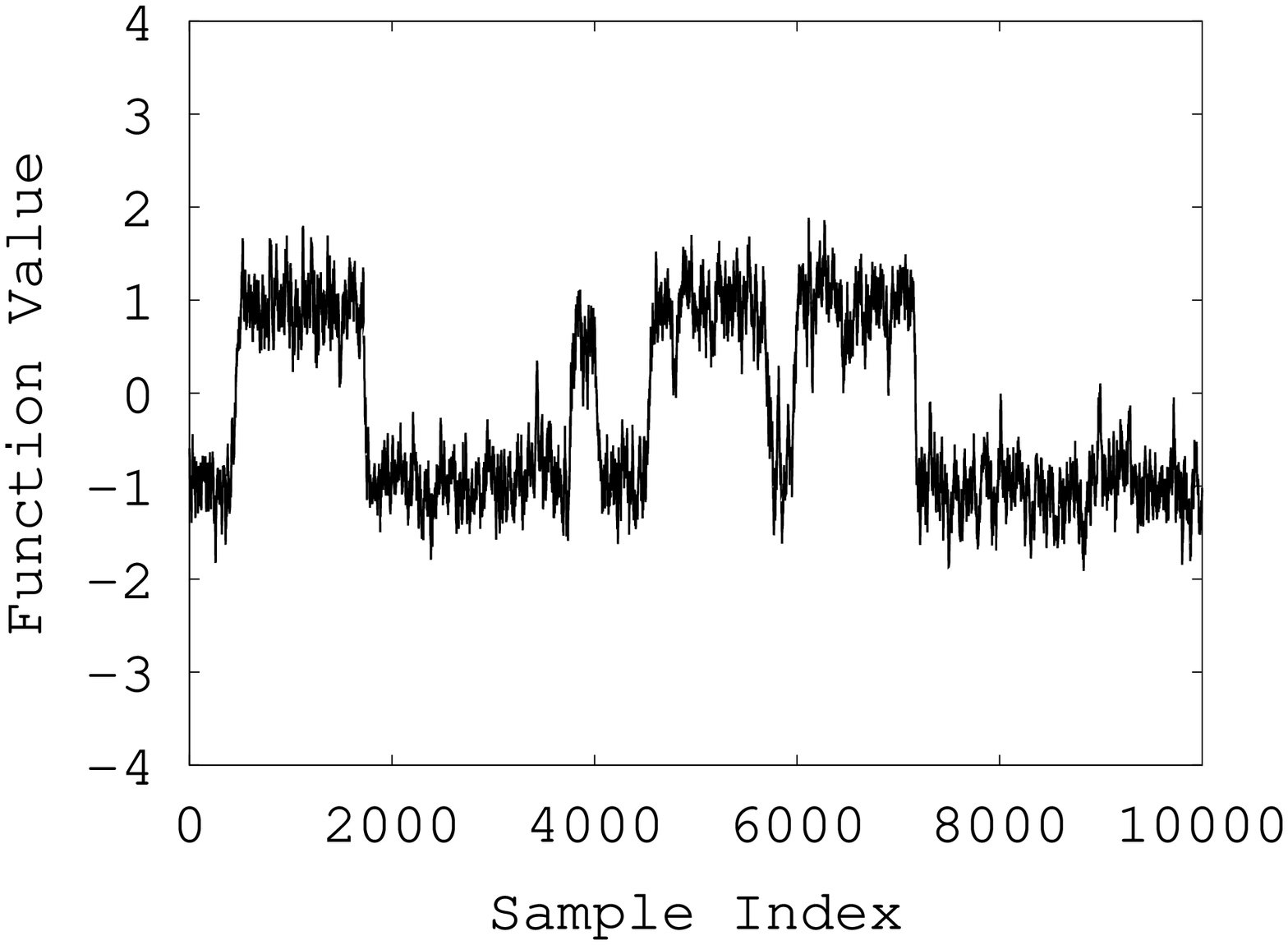}}
\caption{The resulting prediction and the trajectories of observables
in the one-node neural network problem.}
  \label{fig:nn1n_traj}
\end{figure*}

Consider the simplest Bayesian neural network regression model~\cite{Bish96},
having a single node in the hidden layer:
\[y \approx u(\bfq; x) = q_3\tanh(q_1 x + q_2) + q_4,\]
where $(x,y)$ represents a data point. Suppose a total of $100$ data points are given 
as shown by the large dots of Figure \ref{fig:nn1n_traj} (a).
The posterior distribution is
\[-\log\rho(\bfq) =\frac12\beta\sum_{i=1}^{100}(y_i - u(\bfq; x_i))^2
 +\frac12\alpha\|\bfq\|^2 +\mbox{const},\]
 where $\beta = 2.5$ and $\alpha = 0.8$. Suppose the sampler is discretized Brownian dynamics with 
 $\Delta t = 0.01$, and let the Markov chain be obtained by subsampling the original trajectory 
  at time interval $0.1$ (every 10 points). The given observables for finding the maximizing observable
  are the parameters of the model $q_1$, $q_2$, $q_3$ and $q_4$. 
  The prediction for the first data point 
  \[\bar{u}(x_1):=\frac1N \sum_{i=1}^N u(\bfq_i;x_1)\] 
  is also examined as an example of an observable of interest.

Figure \ref{fig:nn1n_traj} (a) shows the overall prediction $\bar{u}(x)$ resulting from the regression model,
 and Figures \ref{fig:nn1n_traj} (b) and (c) show the trajectories of
observables. It can be seen that when the sampling is insufficient, the prediction is inaccurate
(the asymmetry implies only one mode is explored). It can also be seen that the maximizing observable
makes many fewer transitions than $q_1$. This is because the maximizing observable tends to make 
the ``hardest'' moves, hence has longest autocorrelation time.

\begin{figure*}
  \centering
  \subfloat[$\tau$]{\includegraphics[width=0.3\textwidth]{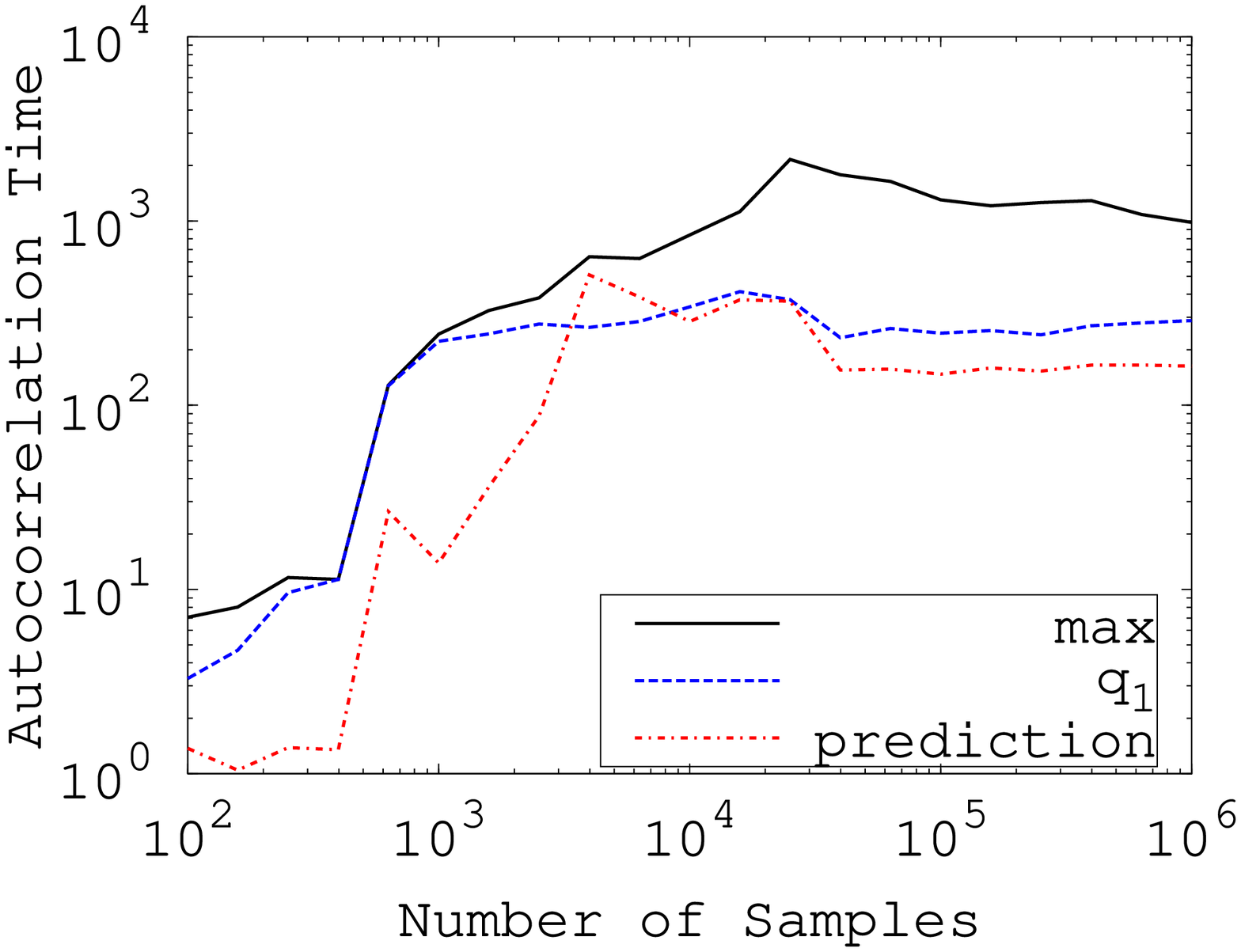}}
  \subfloat[ESS]{\includegraphics[width=0.3\textwidth]{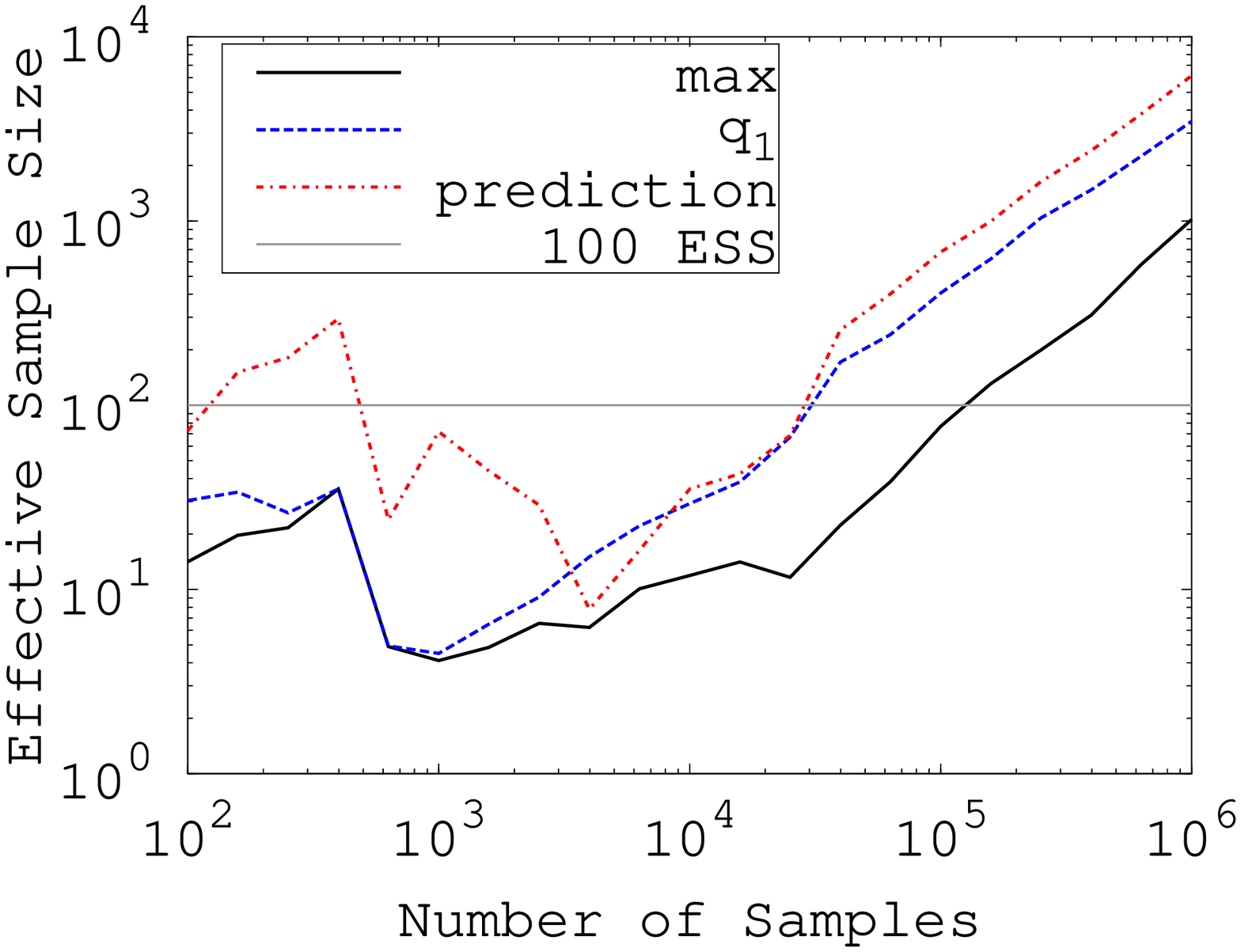}}
  \subfloat[MSE]{\includegraphics[width=0.3\textwidth]{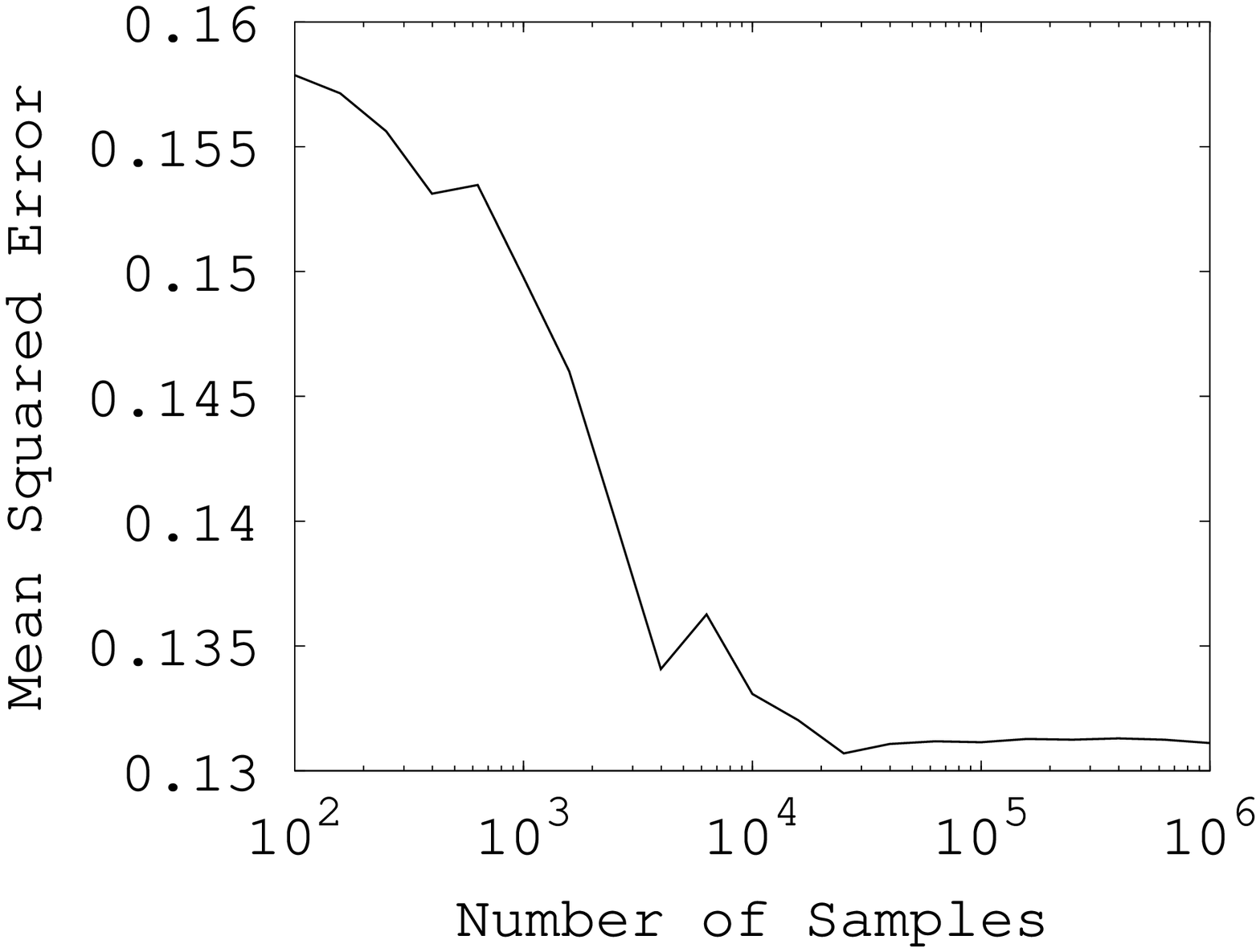}}
\caption{Autocorrelation times, effective sample sizes, and the mean squared error
for the one-node neural network model.}
  \label{fig:nn1n_tauall}
\end{figure*}

Figure \ref{fig:nn1n_tauall} (a) and (b)
show the ESS and $\tau$ for single observables and the maximizing observable.
It can be seen that the ESS of the prediction $\bar{u}(x_1)$ is obviously misleading
for convergence detection, since it indicates sufficient sampling when $N\approx100$.
It can also be seen
that the maximum $\tau$ is significantly larger than the longest $\tau$ of individual
observables. This is a result of combining 4 observables linearly.

Figure \ref{fig:nn1n_tauall} (c) shows the mean squared error of the regression converging when $N>2.0\times 10^4$.
This coincides with the convergence of $\tau_{\max}$. On the other hand, the the $\tau$ for the prediction might suggest stopping far too early
and the $\tau$ for $q_1$ might also be too optimistic.

\subsection{Logistic regression}

Consider a Bayesian logistic regression model~\cite{Gel04}. 
The logistic function maps a linear combination of features $\bfx$ to a probability:
 \[\sigma(\bfq; \bfx) = 1/(1+\exp(-\bfq\T \bfx)). \]
The posterior distribution is
\begin{align*}
-\log\rho(\bfq) =&\beta\sum_{i=1}^{n}(y_i\log(\sigma_i) + (1-y_i)\log(1-\sigma_i))
 \\&+\frac12\alpha\|\bfq\|^2 +\mbox{const.},
 \end{align*}
where $y_i$ is the class label of data example $i$, $\beta = 1.0$, $\alpha = 0.1$, 
and $n$ is the total number of data examples.
When predicting the label of a given data example $\bfx$, use the average value
\[\bar{\sigma}(\bfx)=\frac1N \sum_{i=1}^{N}\sigma(\bfq_i; \bfx)\]
over all samples. The label $t$ of a data example $\bfx$ is predicted to be $1$ if $\bar{\sigma}(\bfx)\ge0.5$
and $0$ if $\bar{\sigma}(\bfx)<0.5$.
For data, use the Australian Credit Approval dataset from the UCI machine learning
data repository \cite{Lich13}. It contains 690 data examples and provides 15 features for each data example. 
Half of the examples in the dataset are extracted to form the training set, and the rest of them
are in the testing set. The sampler is Brownian dynamics with $\Delta t = 0.05$. This step size best balances computation cost
and sampling accuracy (measured by the training error).
As for the one-node neural network problem,
the model parameters $q_i$
are used as observables for estimating $\tau_{\max}$
and the prediction for the first data point in the training set as an observable of interest.

\begin{figure*}
  \centering
  \subfloat[$\tau$]{\includegraphics[width=0.3\textwidth]{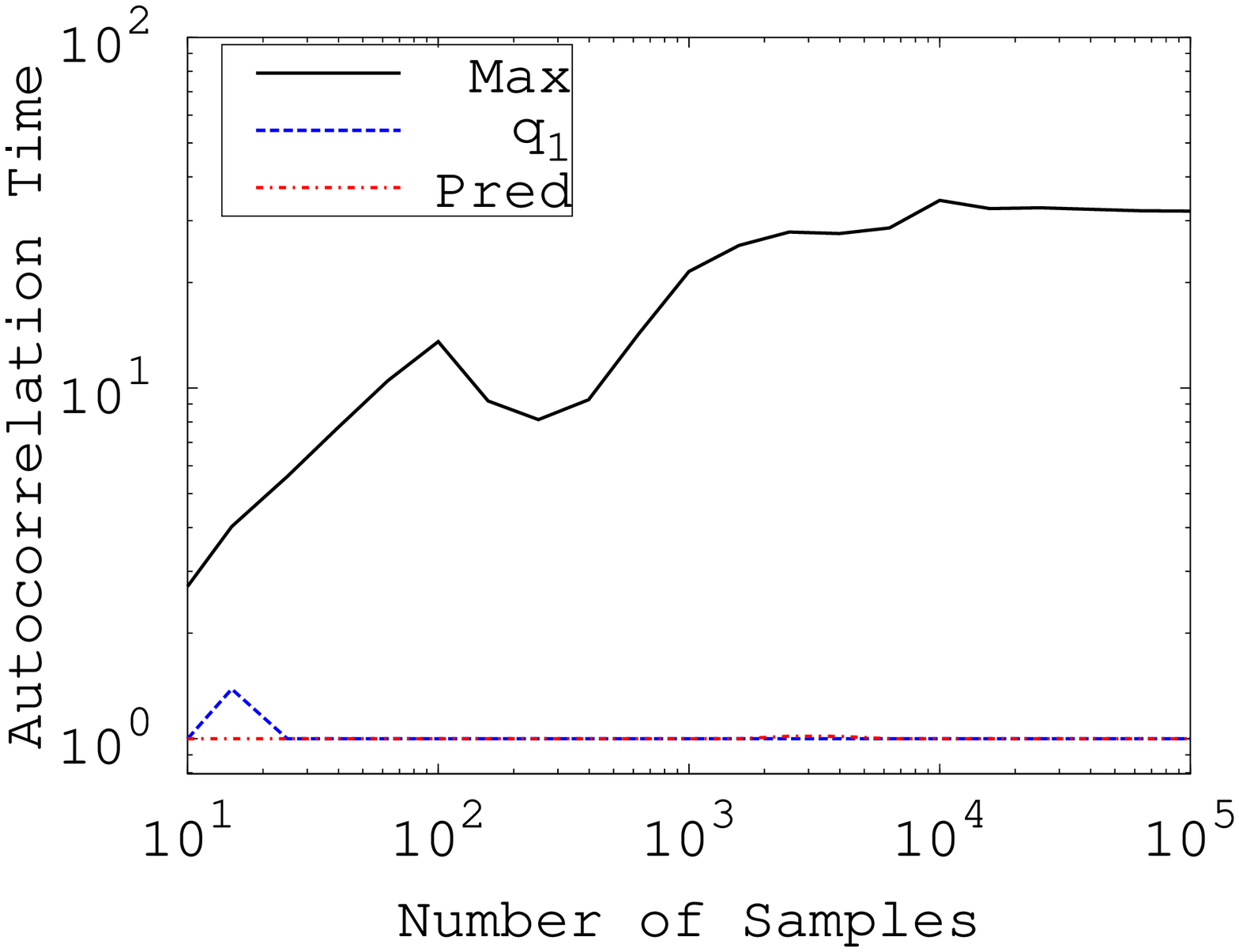}}
  \subfloat[ESS]{\includegraphics[width=0.3\textwidth]{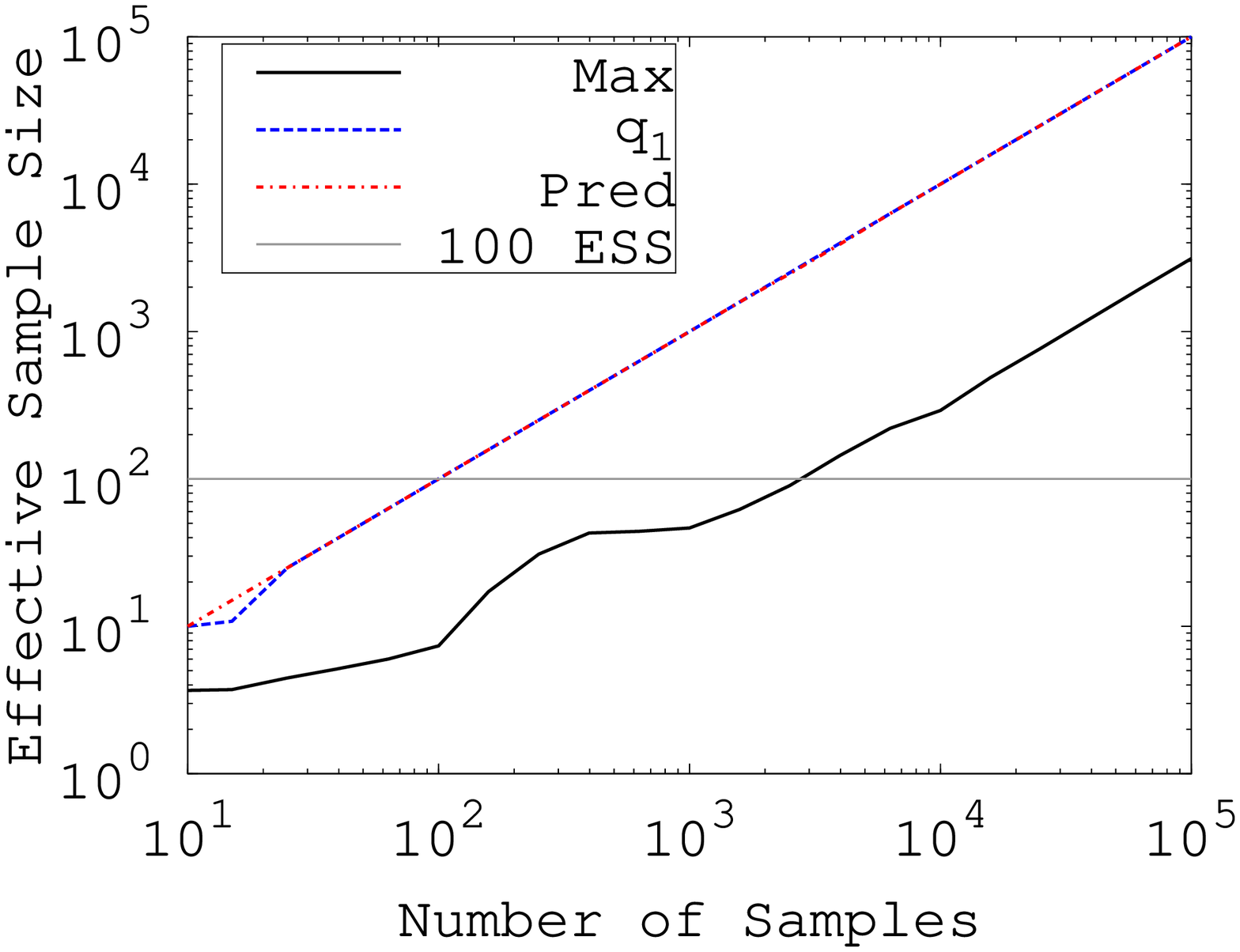} }
  \subfloat[Training error]{\includegraphics[width=0.3\textwidth]{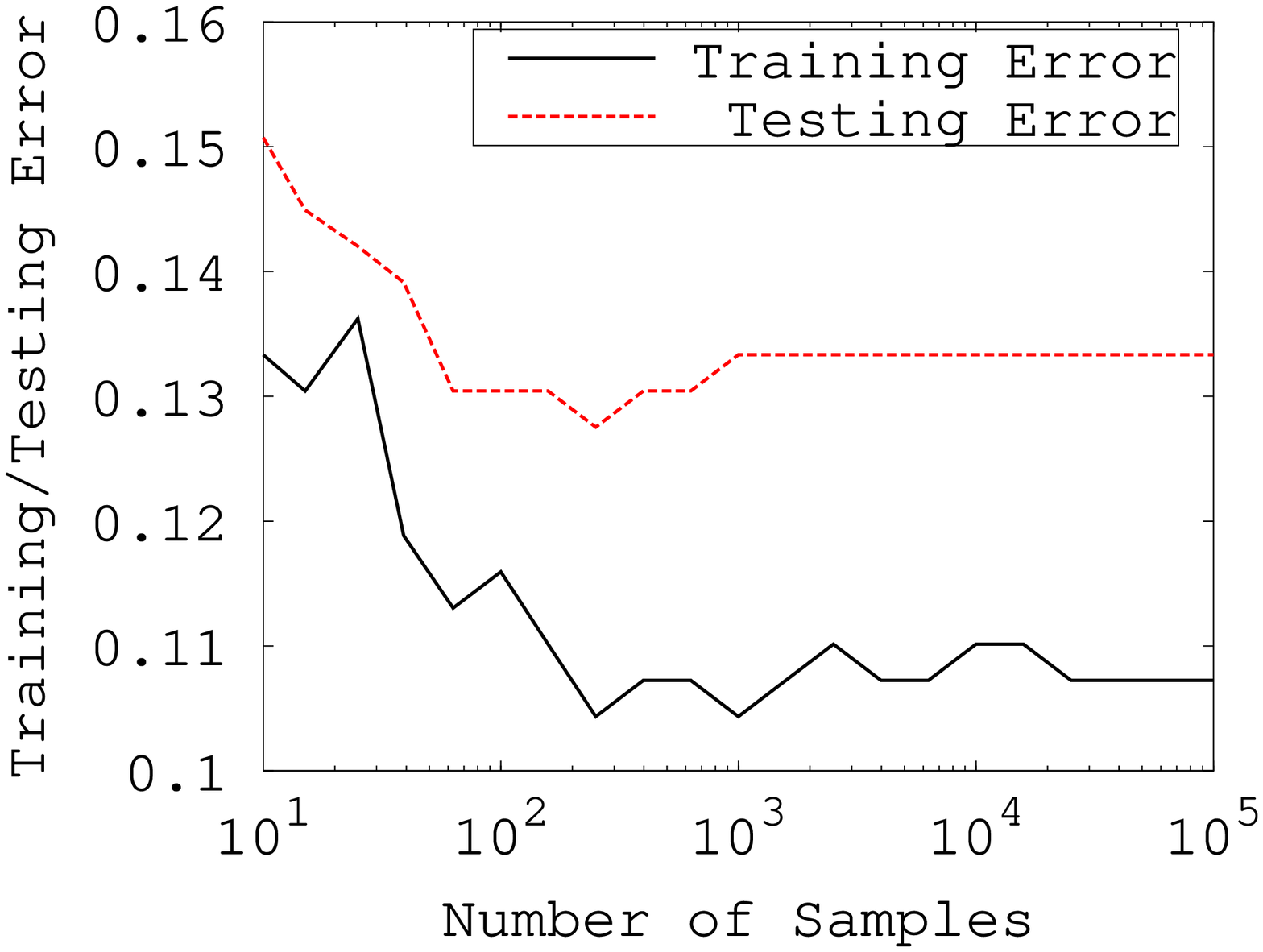}}
\caption{The autocorrelation times, the effective sample sizes,  and the training/testing error
in the logistic regression problem.}
  \label{fig:logistic_tauall}
\end{figure*}

Figure \ref{fig:logistic_tauall} shows $\tau$, the ESS, and the training/testing error in the logistic regression problem.
The training error is defined as
\[err_{\mathrm{train}}=\frac1n \sum_{i=1}^n 1_{t(\bfx_i)\neq y_i}\]
The testing error is defined similarly except the data examples are from the testing set.

In practice, one may monitor convergence by observing the convergence of the training error.
 This approach, however, can be very expensive for large data sets. One has to evaluate the prediction
 function for all data examples. Instead, one might observe the convergence of observables such as the
 prediction function of a single data example or a single weight. But as is shown in the figure, it is risky
 to only observe one of them. Both the training and testing error do not converge 
 until the number of samples exceeds $1000$,
 but both the prediction and the weight$q_1$ have an almost constant autocorrelation time equal to 1 (implying independent samples). This is obviously misleading. On the other hand, 
 the maximizing observable does not show convergence until $N>1000$. This demonstrates that the maximzing observable
 is a more reliable indicator of convergence.

\section{Discussion and conclusions}
There is considerable agreement on the importance
of estimating the integrated autocorrelation time $\tau$.
It is illustrated here and acknowledged by others
that better estimates of $\tau$ are needed.
Derived and presented here are computational methods
based on estimating the greatest possible $\tau$,
which for some problems give more reliable estimates at a cost
that is only a small fraction of that required to generate the samples.
The approach suggested here is quite general,
and there are opportunities for refinement, 
for example, using better basis functions.

\bibliography{16FaCS_submit}
\bibliographystyle{abbrv}

\appendix

\section{Marginally Better Estimates of Integrated Autocorrelation Time}  \label{app:tau}

\subsection{An optimal lag window}  \label{ss:lag}

The autocorrelation function can be highly oscillatory,
which would invalidate the method proposed here.
To obtain a much more slowly varying function,
use values based on doubling: $v_i=u(\bfq_{2i})+u(\bfq_{2i+1})$.
In the case of a reversible sampler,
this can be shown to yield a positive decreasing convex
autocorrelation function~\cite[Sec.~3.3]{Geye92}.
It is straightforward to show that
the autocorrelation time of $u(\bfq)$ can be recovered from that
of $v$:
\[\tau_u = \frac12 \frac{C_v(0)}{C_u(0)}\tau_v.\]

\subsubsection{Form of lag window}

Assume the use of a lag window with
$0\le w(k)\le 1$ and $w(k)$ nonincreasing.
Assume that covariance estimates $C_N(k)$ satisfy
\[ C_N(k) = C(k) +\sigma C(0)\eta_k\]
for some nonnegative value $\sigma$
where the $\eta_k$ are independent standard normally distributed random values.
The estimated autocorrelation time is
\[\tau^\est = 1
 + 2\sum_{k=1}^{+\infty}w(k)\frac{c(k) +\sigma\eta_k}{1 +\sigma\eta_0}\]
where $c(k) = C(k)/C(0)$ is the ACF
and $w(k) = 0$ for $k\ge M$.
Assuming $\sigma\ll 1$,
the error in the integrated autocorrelation time, to a first approximation, is
\[- 2 R
 + 2\sigma\sum_{k=1}^{+\infty}w(k)\eta_k
 - 2\sigma\eta_0\sum_{k=1}^{+\infty}w(k)c(k),\]
 where
 \[R =\sum_{k=1}^{+\infty}(1 - w(k))c(k). \]
The expectation of the square of the error is
\[4 R^2 + 4\sigma^2\sum_{k=1}^{+\infty}w(k)^2
 + 4\sigma^2\left(\sum_{k=1}^{+\infty}w(k)c(k)\right)^2,\]
which is minimized
for a given value of $m$
if
\begin{equation}  \label{eq:wk}
w(k) =\frac{c(k)}{\sigma^2}\left(R -\sigma^2\sum_{k=1}^{+\infty}w(k)c(k)\right).
\end{equation}
Note that, for small enough $\sigma$,
the right-hand side is positive.

The ACF $c(k)$ is unknown,
so,
{\em for the purpose of defining the lag window},
it is suggested to
use the specific model
\[ C_N(k) = c_0\lambda^k +\sigma c_0\eta_k,\]
where $\lambda$, $c_0$, and $\sigma$ are determined as
in Section~\ref{sss:fit} that follows.
The goal is to capture the behavior of the tail of the ACF,
which will be achieved by using enough points $M$ in the fitting.
Then
$c(k) =\lambda^k$,
so
$w(k)$ is proportional to $\lambda^k$, suggesting the choice
\begin{equation}  \label{eq:wka}
w(k) =\min\{1,\lambda^{k-m}\}
\end{equation}
where $m$ is to be determined optimally.

Using the specific model for $C_N(k)$,
the expectation of the error squared becomes
\[\frac{4c_0\lambda^2}{(1 -\lambda^2)^2}
  \left(\mu^2 +\sigma^2(1 +\lambda-\mu)^2\right)
 +\frac{4\sigma^2\log\mu}{\log\lambda} -\frac{4\sigma^2}{1 -\lambda^2}\]
where $\mu =\lambda^m$.
Differentiating with respect to $\mu$ and setting the result to zero
leads to a quadratic equation for $\mu$,
which can be shown to have a positive and negative root.
If the positive root exceeds 1,
this indicates that the sample size $N$ is not large enough.

\subsubsection{Fitting the autocorrelation function}  \label{sss:fit}
With residuals
\[r_k = c_0\lambda^k - C_N(k),\]
the parameters $c_0$ and $0<\lambda < 1$ minimize
$\sum_{k=0}^{M-1} r_k^2$.
The value of $c_0$ that minimizes this must satisfy
\[\sum_{k=0}^{M-1}\lambda^k r_k = 0,\]
whence $c_0 ={Q(\lambda)}/{P(\lambda)}$,
where
\[P(\lambda) =\sum_{k=0}^{M-1}\lambda^{2k}\mbox{ and }
Q(\lambda) =\sum_{k=0}^{M-1}C_N(k)\lambda^{k}.\]
Substituting this into the objective function gives
\begin{equation}  \label{eq:obj}
\sum_{k=0}^{M-1} r_k^2
 = -\frac{Q(\lambda)^2}{P(\lambda)} +\sum_{k=0}^{M-1} c(k)^2.
\end{equation}
The optimal value of $\lambda$
is obtained by minimizing this rational function.
(One can use golden section search beginning with
the interval $[0, 1]$.
We used bisection applied to the gradient.)
The ML estimate of the standard deviation $\sigma$ of the noise
is obtained from the RMS of the resulting residual.

One can simply use the exponential model
for the tail $\sum_{k=M}^{N-1}w(k)C_N(k)$ of $\tau^\mathrm{est}$.

\subsection{A comparison with the lag window of {\tt acor}}

\begin{figure*}
  \centering
   \subfloat[$H_1$]{\includegraphics[width=0.3\textwidth]{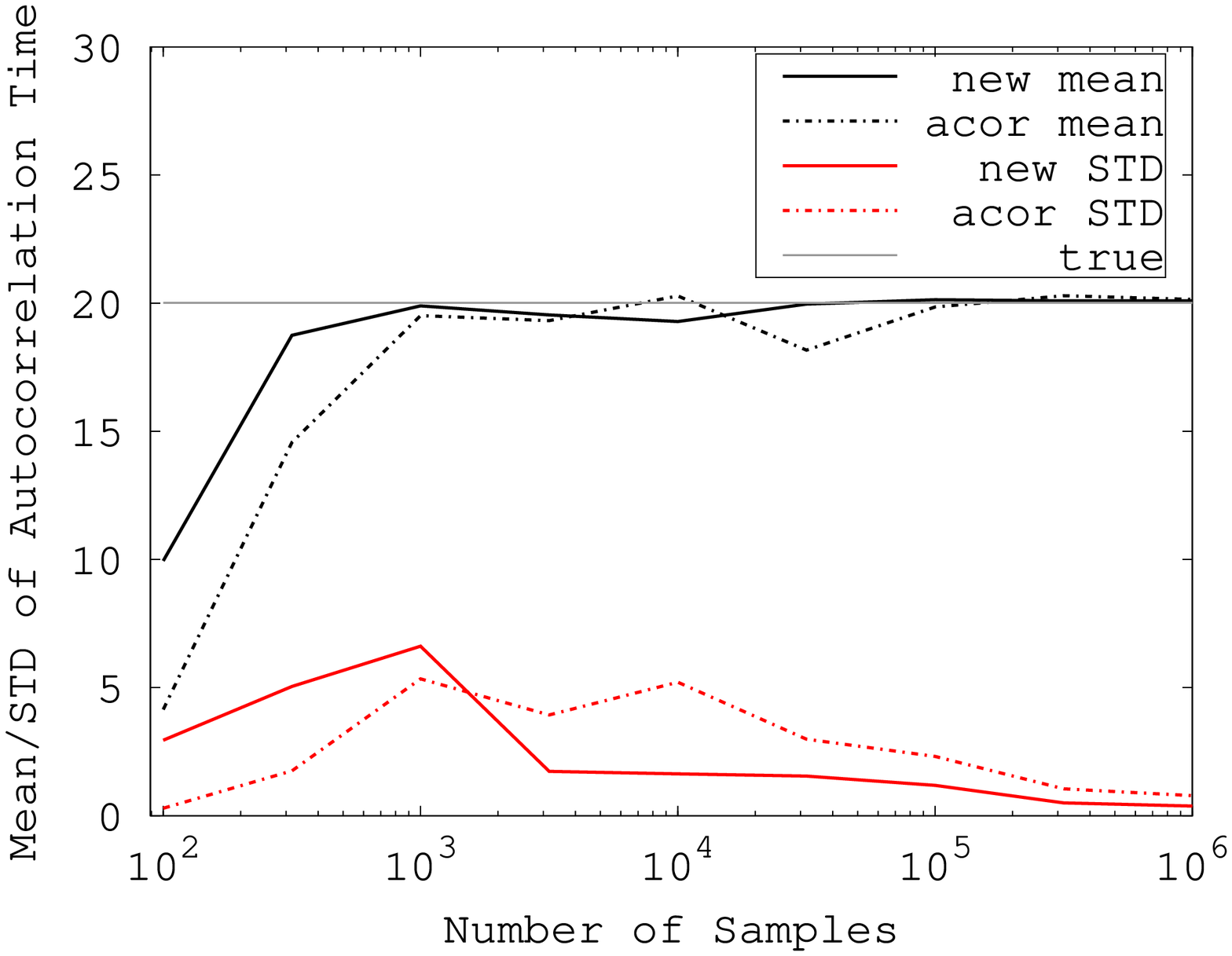}}
   \subfloat[$H_1+H_2+H_3$]{\includegraphics[width=0.3\textwidth]{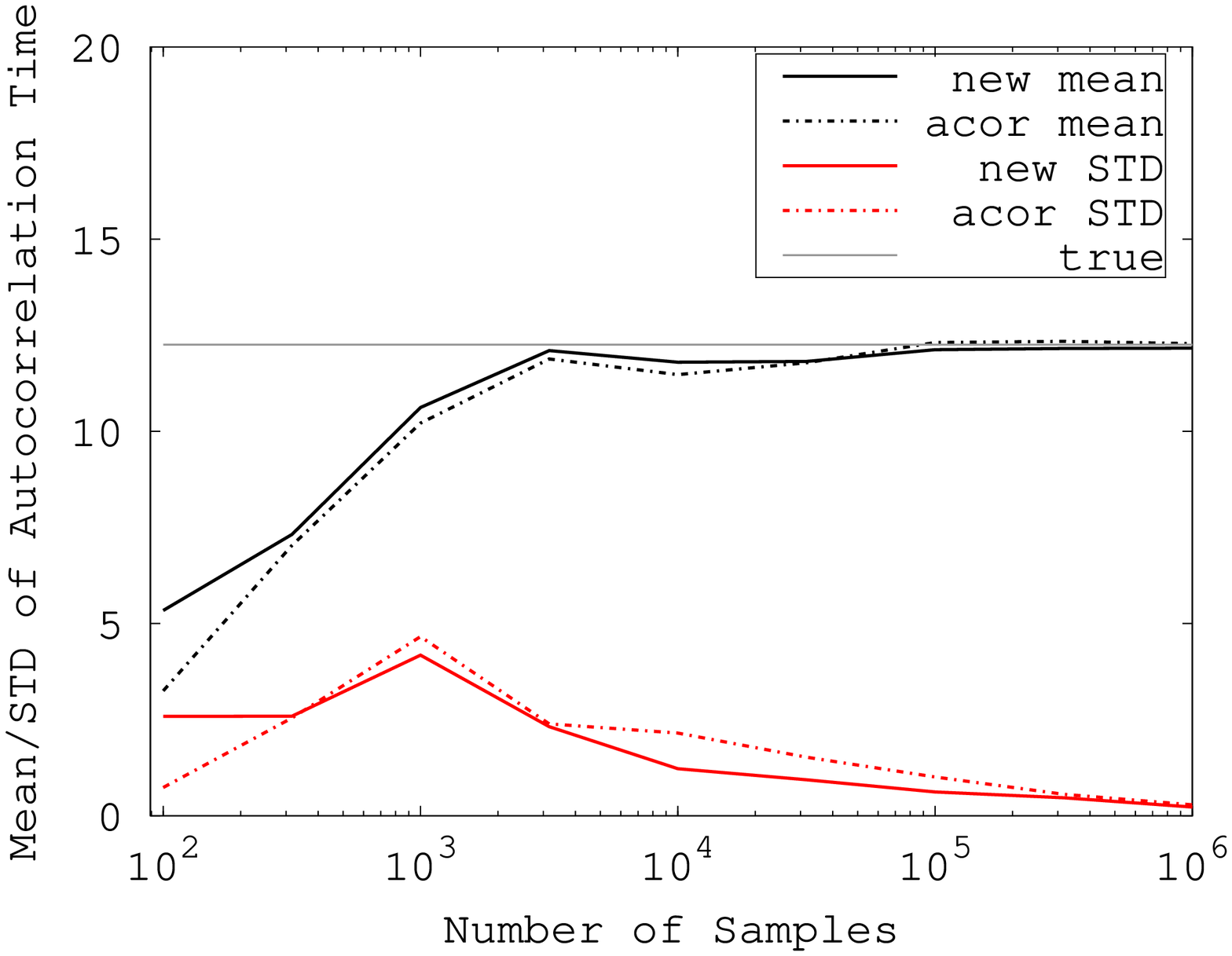}}
   \subfloat[$\tau_{\max}$]{\includegraphics[width=0.3\textwidth]{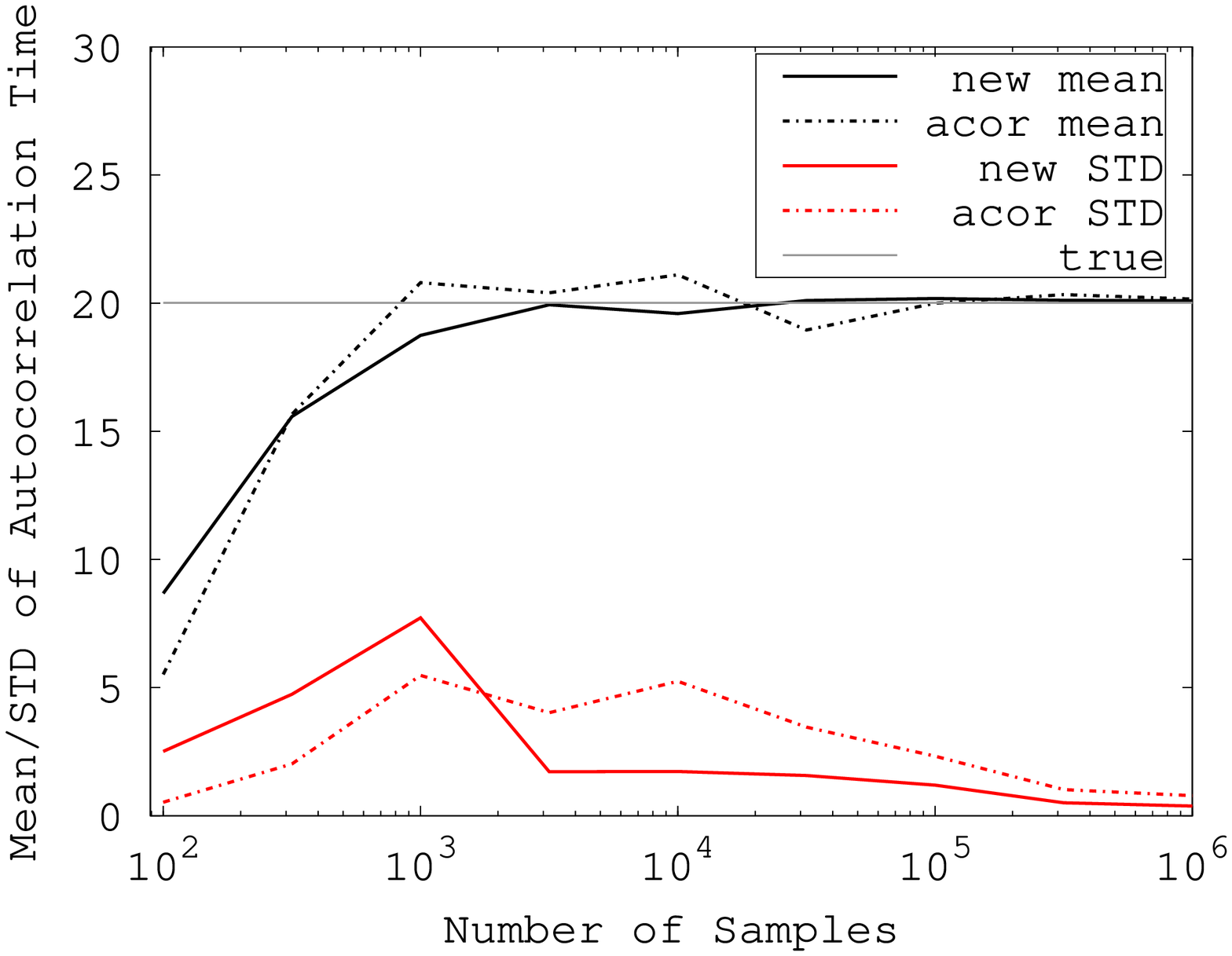}}
  \caption{Estimated $\tau$ and $\tau_{\max}$ by the new lag window and {\tt acor} for the 1D 
  standard Gaussian.}
  \label{fig:1dgauss_tauall_compare}
\end{figure*}

Figure \ref{fig:1dgauss_tauall_compare} shows the 
estimated $\tau$'s of $H_1$ and $H_1+H_2+H_3$ as well as $\tau_{\max}$. 
 It can be seen that when $N$ is small,
both methods underestimate $\tau$, but the new one gives larger estimates.
When $N$ is large, both methods converge to the true value, 
and the new method has a smaller variance.
Overall, the proposed lag window is more reliable than acor.

\subsection{The lag window for extremely small ESS}
\begin{figure*}
  \centering
  \includegraphics[width=0.75\textwidth]{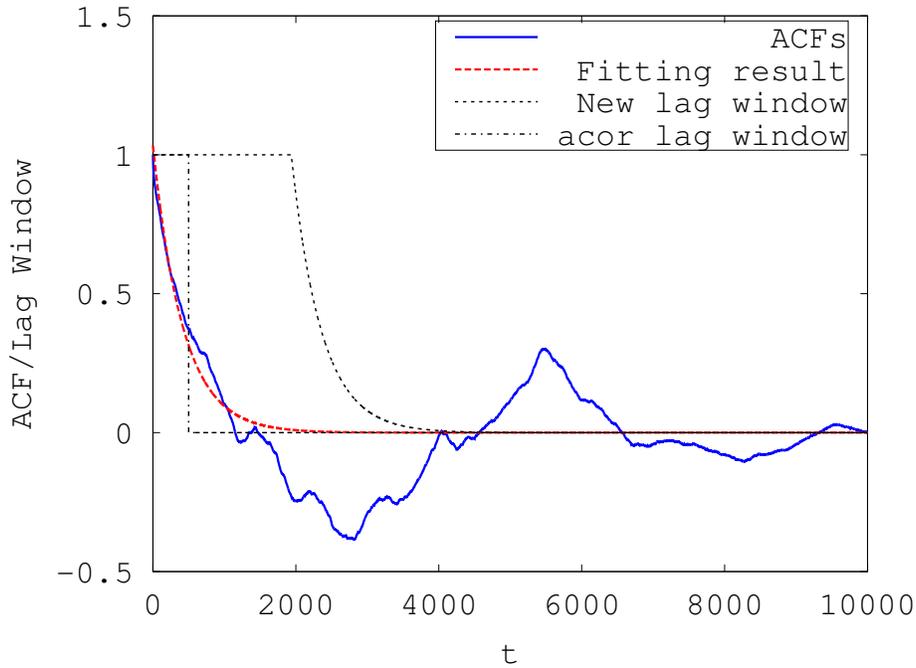}
  \caption{An example of ACF and lag window for the one-node neural network problem.}
  \label{fig:nn1n_m}
\end{figure*}

It might be difficult to get a good lag window when the number of samples
is extremely insufficient.
Figure \ref{fig:nn1n_m}
shows the lag window obtained  by the new method and {\tt acor} in the one-node neural network problem.  
For {\tt acor}, the lag window obviously contains not enough data. And for the proposed method, it contains perhaps
too much data since some of the data is already in the negative value range due to statistical error. The noise
for the auto-correlation function estimator is highly correlated when $\tau$ is large. In this example, the effective
sample size is less than 10. In such cases, $\tau$ may be underestimated by both methods. 

\end{document}